\documentstyle[emulateapj,epsfig]{article}

\newcommand{\kms}{km~s$^{-1}$}
\newcommand{\Msun}{M_\odot}
\newcommand{\be}[1]{\begin{equation}\label{#1}}
\newcommand{\ee}{\end{equation}}
\newcommand{\ttau}{\mbox{\boldmath $\tau$}}

\begin{document}
\title{Velocity Field Statistics in Star-Forming
Regions. I. Centroid Velocity Observations}

\author{Mark S. Miesch\footnote{Now at DAMTP, 
University of Cambridge, Silver Street, Cambridge, CB3 9EW, UK 
; M.S.Miesch@damtp.cam.ac.uk}}
\affil{Laboratory for Astronomy and Solar Physics \\
NASA GSFC, Code 682 \\
Greenbelt, MD \\ miesch@ktaadn.gsfc.nasa.gov}

\author{John Scalo,}
\affil{Department of Astronomy \\
University of Texas at Austin
Austin, TX 78712 \\ parrot@astro.as.utexas.edu}

\and

\author{John Bally}
\affil{Center for Astrophysics and Space Astronomy \\
University of Colorado, 
Boulder, Colorado 80309-0389 \\ bally@casa.colorado.edu}

\slugcomment{Draft date: \today}

\lefthead{Miesch, Scalo \& Bally}
\righthead{Velocity Field Statistics in SF Regions}
\begin{abstract}
	The probability density functions (pdfs) of molecular line centroid
velocity fluctuations, and of line centroid velocity fluctuation
differences at different spatial lags, are estimated for several 
nearby molecular clouds with active internal star formation.
The data consist of over 75,000 $^{13}$CO line
profiles divided among twelve spatially and/or kinematically distinct
regions.  These regions range in size from less than to 1 more than 40pc
and are all substantially supersonic, with centroid fluctuation Mach numbers 
ranging from about 1.5 to 7.  The centroid pdfs are constructed using 
three different types of estimators.  Although three regions (all in 
Mon R2) exhibit nearly Gaussian centroid pdfs, the other regions show 
strong evidence for non-Gaussian pdfs, often nearly exponential, with 
possible evidence for power law contributions in the far tails.  
Evidence for nearly exponential centroid pdfs in the neutral HI component 
of the ISM is also presented, based on older published data for optical 
absorption lines and HI emission and absorption lines.  These
strongly non-Gaussian pdfs disagree with
the nearly-Gaussian behaviour found for incompressible turbulence (except
possibly shear flow turbulence) and
simulations of decaying mildly supersonic turbulence.  Spatial images of
the largest-magnitude centroid velocity differences for the star-forming
regions appear less
filamentary than predicted by decay simulations dominated by vortical 
interactions.  No evidence for the scaling of difference pdf kurtosis 
with Reynolds number, as found in incompressible turbulence experiments 
and simulations, is found.  We conclude that turbulence in both 
star-forming molecular clouds
and diffuse HI regions involves physical processes which are not adequately 
captured by incompressible turbulence or by mildly supersonic decay 
simulations.  The variation with lag of the variance and kurtosis 
of the difference pdfs is presented as a constraint on future simulations, 
and we evaluate and discuss the implications of the large scale and 
Taylor scale Reynolds numbers for the regions studied here.
\end{abstract}
\keywords{ISM: clouds, ISM: molecules, ISM: kinematics and dynamics, 
stars: formation, turbulence}

\section{Introduction}
         Although a great deal of effort has been devoted to quantitatively
describing the complex column density spatial structure of star-forming
regions (for
recent approaches see \cite{falga90} 1990, 1991; \cite{gill90} 1990;
\cite{lange93} 1993; \cite{adams94} 1994; \cite{wisem94} 1994; 
\cite{zimme93} 1993; \cite{houla92} 1992; \cite{scalo90} 1990; 
\cite{chapp98} 1998; \cite{stutz98} 1998), comparatively
little attention has been paid to characterising the radial velocity
dimension of the data.  Exceptions are
studies of possible velocity dispersion-size scaling relations (see
\cite{falga92} 1992, \cite{barra98} 1998, \cite{goodm98} 1998, and references
given there),
 estimation of the velocity correlation
function and related 2-point statistics (see \cite{hobso92} 1992; 
\cite{kitam93} 1993; \cite{miesc94a} 1994 and references to earlier work given
there) and searches for evidence of rotation (e.g. \cite{goodm93} 1993).
Since one expects a signature of the dynamical and physical processes
to appear in the velocity field, and because the velocity field is strongly
coupled to, and may in a sense control,
the density field and even the star formation rate and the IMF (see models
by, e.g. \cite{fleck83} 1983, \cite{silk95} 1995, 
\cite{scalo98b} 1998)  it is important to develop additional diagnostics to
investigate it.   As a step toward a
better understanding of molecular cloud velocity structure, Falgarone and
coworkers (\cite{falga89} 1989, \cite{falga91b} 1990, 1991, 
\cite{falga94} 1994; see below
for discussion) have explicitly tried to relate radial velocity
information to dynamical processes through the comparison of observed line
profiles with frequency distributions, or probability distribution
functions, found in experimental and simulation studies of turbulence. 
In an earlier short report we (\cite{miesc95} 1995)
extended that program to the frequency distribution of
line centroid velocities.  In the present paper we give a more detailed
description and discussion of both the
observational data and the centroid probability density functions (which we
hereafter refer to as centroid pdfs), and further extend the study to the
pdfs of centroid velocity differences.  Pdfs of line widths and
line skewnesses are examined in a separate paper 
(\cite{miesc99} 1999, hereafter Paper II).

The primary goal of this work is to provide useful quantitative
observational constraints on ideas, models, and simulations of 
interstellar turbulence and its relation to star formation--constraints 
which must be accounted for by any theoretical approach that purports to 
provide a physical explanatory understanding of the phenomenon.
A recurring theme in the present paper is the degree to which interstellar 
turbulence resembles incompressible turbulence, a field
with a very large experimental and theoretical literature.  
However, we think it is important to recognise from the
outset that even if interstellar turbulence {\it does} turn out to
resemble incompressible turbulence in some respects,
this would not imply that interstellar turbulence is ``understood.''
Incompressible turbulence remains an essentially
unsolved problem.  As many authors have pointed out, there has been
(arguably) little tangible progress in the field of
incompressible turbulence since Kolmogorov's seminal paper (\cite{kolmo41}
1941, K41), and most of it consists of
elucidating the ways in which K41 was incorrect (e.g. the variation of the
scaling exponents of different-order
structure functions), and a proliferation of contrasting phenomenological
theoretical approaches, mostly imported from
other branches of physics (e.g. statistical mechanical and field
theoretical approaches; see \cite{she97} 1997 and \cite{lvov97} 1997).  
If interstellar turbulence resembles incompressible turbulence, this result 
may help exclude some physical processes as being dominant, and may
implicate the nonlinear advection operator in the momentum equation (which
controls incompressible turbulence) as the
dominant physical process.  However, the physics of this operator in the
presence of a large number of degrees of freedom
(i.e. large Reynolds number) is still obscure, and one must not confuse the
existence of a huge literature with
understanding.  Furthermore, compressible turbulence does not
entail any quadratic invariants (e.g. 
quantities conserved by the advection operator
which are quadratic in the velocity, like kinetic energy), a
property which is central to all models of
incompressible turbulence. This should from the outset lead one to expect that
interstellar supersonic turbulence is
different in fundamental ways from incompressible turbulence.  From this
point of view, the present work is aimed at
uncovering observationally these differences.

In \S\ref{review} we present a discussion of the advantages and disadvantages
of the use of centroid pdfs instead of line profiles, and a summary of
previous
work using this approach (and the velocity difference pdf) for incompressible
turbulence, molecular interstellar regions, and extragalactic structure.
We also
present centroid velocity pdfs culled from early surveys of
optical absorption lines and HI emission
and absorption lines in order to show that good evidence already exists
for roughly exponential centroid velocity
pdfs in the lower-density atomic HI component of the ISM.  The
data employed in the present study of molecular star-forming regions
are presented and discussed in \S3, and an overview of the
dynamics, star formation activity, and physical environment in each region
is provided in Appendix \ref{overview}.
In \S\ref{results} the statistical results are presented, including centroid velocity
images and pdfs, as well as velocity difference images, pdfs and pdf moments.
Several parametric and non-parametric pdf estimators are used and compared.
For the velocity differences, the variation of the pdfs with lag is emphasised
and quantified and the possible relation of the non-Gaussian tails to
filamentary structures is investigated.  The centroid velocity and velocity
difference maps are also used to derive effective Reynolds numbers and Taylor
scales for the regions considered.  The results are summarised
in \S\ref{summary}.  This is the first in a series of papers on the statistical analysis
of velocity fields in star-forming regions.  In Paper II 
(\cite{miesc99} 1999),
we use the same data sets presented here to investigate spectral linewidth
and line skewness variations, which provide yet another important diagnostic of 
molecular cloud velocity fields available from densely-sampled emission line
observations.  
In the final paper of the series (\cite{scalo99} 1999, hereafter
Paper III), we provide theoretical interpretations of the observed centroid 
velocity, velocity difference, and linewidth pdfs and a comparison to analytic 
and numerical models.

\section{Velocity pdfs and their Estimation}\label{review}
\subsection{General Background}\label{background}

The one-point probability density function contains information that
is qualitatively different from correlation functions and related
two-point statistics, which are {\it moments} of some probability density.
For example, the autocorrelation function or structure function,
while containing spatial information, basically only involve the {\it
variance}
of the two-point probability density of velocities as a
function of scale.  The two-point probability density function itself could
be computed, but is difficult to
visualise, since, for a given lag scale, it is a two-dimensional surface,
and its full representation would involve
such surfaces at different spatial lags.  More information about the
velocity two-point pdf would require
high-order moments, which cannot be constructed for ISM data because of
noise.  In contrast, the one-point velocity
pdf, although it contains no spatial information, is easily displayed (it
is basically just a one-dimensional
histogram), giving access to the probability structure of the velocity
field.  Similarly, correlations between
linewidth and other physical parameters only involve the variance of the
one-point pdf, and thus averages over the information in the one-point pdf
itself.  For these reasons, the
one-point velocity pdf offers a qualitatively different view of the
velocity field.

The relative independence of the velocity pdf and a second order moment of
the two-point pdf, like the power spectrum or
correlation function, can be seen in the work of \cite{dubin95} (1995), 
who generated simulated line
profiles from random velocity fields with a prescribed power spectrum.
Although they emphasised the ability of a
Kolmogorov energy spectrum (k$^{-5/3}$) to yield non-Gaussian
line-profiles, their results show that other forms of the
energy spectrum would yield similar results.  In particular, 
a steeper k$^{-2}$ spectrum, which might be expected for a
field of discontinuities or shocks, also gave non-Gaussian profiles
of similar form.  Thus the 2-point velocity correlation function (a
moment) is probably only weakly coupled to the 1-point velocity pdf.

 Recent work in several areas suggests that the one-point
probability distribution function of
dynamical variables like velocity is a useful tool that may be sensitive to
dynamical processes.  These studies include large scale structure of galaxy
velocities (\cite{berna94} 1994;  \cite{kofma94} 1994;
\cite{catel94a} 1994a,b),
incompressible terrestrial turbulence (see below),  distinguishing
nonlinear chaotic processes from stochastic processes (\cite{wrigh93} 1993), 
and characterisation of samples of musical volume fluctuations
(\cite{scalo98} 1998).  In particular, studies of incompressible
turbulence have shown that the higher moments (skewness, kurtosis,...) of
the pdf can be used to constrain physical models for turbulent
intermittency.  Although non-zero skewness must exist at some level in order 
to provide energy transfer among different scales, the pdf of the velocity 
field itself in incompressible turbulence is in general very nearly Gaussian, 
at least on large enough scales (\cite{batch53} 1953 and \cite{monin71} 1971 
review early work; see more recent experiments and simulations in
\cite{ansel84} 1984, Figure 1; \cite{kida89} 1989, Figure 6;
\cite{jayes91} 1991, Figure 1; \cite{chen93} 1993, Figure 3.).
An important exception is the 3-dimensional incompressible
simulation of homogeneous shear flows by \cite{pumir96} (1996), 
who finds non-Gaussian, nearly exponential, velocity fluctuation
pdfs for velocity components perpendicular to
the streamwise component.  See also \cite{lamba97} (1997)
for non-Gaussian velocity pdfs in channel flow close to the boundary.

 For incompressible turbulence
non-Gaussian behaviour is well-established for velocity
{\em differences} at small scales and velocity
{\em derivatives}, and there is strong evidence from experiments
and simulations for non-Gaussian behaviour in many other variables (see the
papers referred to above and \cite{chen89} 1989; \cite{casta90} 1990;  
\cite{vince91} 1991; \cite{she93} 1993).  Often the pdf of the
velocity difference or derivative field exhibits a near-exponential
behaviour at smaller and smaller scales, and much work has gone into
understanding this behaviour physically, especially in terms of the
stretching properties of the advection operator (see \cite{she91} 1991 
for a review).   Part of the motivation of the present work is to
investigate whether any of these properties occur in the more complex
``turbulence'' of interstellar clouds, and whether even the velocity
fluctuation field itself presents measurable deviations from a Gaussian
pdf.

\cite{falga90} (1990, 1991) have shown that line profiles
constructed from high-sensitivity CO molecular line data (in several
transitions) exhibit excess wing emission, relative to a single Gaussian,
over a very large range of scales, from 0.02 to 450 pc .  For all these
line profiles the width of the wings is about 3 times the width of the line
core if both are fit by Gaussians, but the fractional intensity of the wing
component (fraction of mass at high velocities) varies between about 0.03
and 0.8.   Broad wings were also found in high latitude molecular clouds by
\cite{blitz88} (1988).  The presence of similar broad wings in
regions
whose scales are gravitating and non-self-gravitating, and in regions with and
without internal massive star formation,
suggests that the behaviour is not due to stellar winds or collapse motions,
and the variation
in wing width  in these regions seems to rule out a dilute warm gaseous
component, as pointed out by Falgarone \& Phillips.
Since the line profile, in the optically thin case, is in effect
a histogram of radial velocities, the broad wings have been viewed
in the context of non-Gaussian pdfs, although there is some confusion
concerning whether the line profile should be interpreted as the pdf of
average line of sight
velocities or of velocity differences; the latter interpretation is adopted
by \cite{falga90} (1990) in comparisons with laboratory data.

It is not clear that line profiles give a valid representation of the
velocity pdf.  Every
line profile samples a line-of-sight velocity field which in general
contains a component whose characteristic scale is a significant fraction
of the sample depth.  The
form of these systematic line-of-sight motions is unknown and may severely
limit the correspondence between the line profile and velocity pdf.
Such problems can largely be circumvented in analyses of simulations,
where it is possible to to insure homogeneity on the largest scales
(as in \cite{porte94} 1994, as analysed by \cite{falga94} 1994), 
but homogeneity is probably not a good assumption in general for 
interstellar clouds.  It is not difficult to show that the addition of 
a systematic component can significantly alter the estimate of the 
distribution of the velocity {\em fluctuations}, which is the 
function of interest.  A cloud in non-uniform
rotation about its center, for example, will yield non-Gaussian line
profiles along lines of sight displaced from the projection of the rotation
axis onto the plane of the sky (provided this projection is nonzero).  In
particular, these profiles will exhibit apparent excess wing emission due
solely to the smearing arising from the variation of the
line-of-sight component of the rotational velocity with depth in the cloud,
which will thus distort the pdf of velocity fluctuations.
In addition, radiative transfer effects can distort emission lines and
cause the wings to become relatively more prominent if the cloud is
optically thick (although \cite{falga90} 1990 argue against
this interpretation of the broad wings on the basis of their observed
shapes).

 An alternative procedure, which we adopt here, is to estimate the
pdf of {\em centroid line velocities}  (intensity weighted average velocity
along the line of sight) sampled over a densely observed individual star
formation region.
While a  ``line profile'' is a measure of the radial velocity (or
velocity difference) pdf sampled along the line of sight,
either in a single beam or averaged
over many beams, the ``centroid pdf'' is the pdf of the mean velocity of
line profiles taken over a large spatial sample of positions in the
plane of the sky.  The two functions differ in the direction along which the
sampling for the pdf is taken, and in the quantity sampled.

The advantages of the centroid pdf approach include the much lower
sensitivity required for each of the individual line profiles and the
weaker dependence of the results on large scale systematic motions which,
although still a concern, will tend to be mitigated by the line-of-sight
averaging and by space filtering of the velocity maps (see \S3.1).  For
example, the centroid velocities of the rotating cloud discussed above 
will vary in an obvious way with position, and the effects of rotation can
therefore be removed by applying an appropriate filter.  Such a procedure
is not possible with the individual profiles unless the rotation curve of
the cloud is known. In addition, the presence of a warm ``interclump'' 
medium, or of ``optical depth broadening'', which would both contribute
to the line profiles, will not much affect
the pdf of centroid velocities, since the thermal component and the
line saturation are
symmetric (although the centroid pdf, in the optically thick case,
would only sample fluctuations on the leading edge of the cloud).
The problem with this approach is that the number of
velocities (positions) which must be sampled in order to
accurately estimate the tails of the pdf is very large, at least of order
1000.  Furthermore, the relationship between the pdf of an average
line-of-sight quantity (centroid velocity in this case) and the pdf of the
radial velocity distribution in three dimensions is not clear.

It is also possible to use the centroid velocities to construct the pdf of
velocity {\it differences} for regions
separated by a given distance, or, ``lag'', and examine how this pdf depends
on lag.  These velocity difference pdfs
are commonly used in studies of incompressible turbulence (see references
above), and have been used to study
models for the cosmological evolution of galaxy velocities
(\cite{peebl76} 1976; \cite{ueda93} 1993; \cite{seto98} 1998).  It is
particularly intriguing that the galaxy
velocity difference pdfs exhibit exponential forms at small separations,
very similar to the ISM pdfs reported here.

A preliminary account of the
velocity difference pdfs obtained for the star-forming regions studied here
was presented by \cite{miesc94c} (1994).  Centroid velocity difference pdfs have 
also been reported for CO lines in the $\rho$Oph region (\cite{lis98a} 1998; 
\cite{lis98b} 1998) and for HI lines in the Ursa Major cirrus cloud 
(\cite{mivil98} 1998).  \cite{lis96} (1996, 1998) find that the
velocity difference pdfs in simulations of mildly supersonic decaying 
turbulence exhibit strong non-Gaussian tails which are associated with 
filamentary structures and regions of large vorticity, and they report
some evidence of such behaviour in molecular cloud observations, at least 
in quiescent regions.  Part of the work presented here is aimed at 
estimating the pdfs and spatial distribution of velocity differences 
for regions in which vigorous high-mass star formation has taken place 
(Orion B and Mon R2 regions) although smaller regions containing only 
lower-mass YSOs are also represented (L1228, L1551, HH83).  It should be 
recognised at the outset that the pdfs of the centroid velocities themselves 
for some of these regions already exhibit strongly non-Gaussian tails, 
(\cite{miesc95} 1995; see also \S4.1 below), showing that the 
``turbulence'' in star-forming molecular clouds is different 
from incompressible and mildly supersonic turbulence, which generally 
exhibit nearly Gaussian velocity pdfs.  Apart from 
\cite{miesc95} (1995), we know of only one other study in which 
centroid velocity pdfs of molecular line data have been presented, 
that of \cite{padoa97} (1997) who plotted histograms of centroid 
velocities for several areas in the Perseus cloud, another active
star-forming region.  Although large-scale motions were not filtered 
out and the pdf tails were not investigated in detail, these data 
too clearly exhibit non-Gaussian shapes.

\subsection{Previous Estimates for the Atomic ISM}\label{otherwork}
         Work aimed at determining the pdf of interstellar gas motions dates
back to the early 1950s.  Several studies used optical absorption line
velocities of ``clouds'' along the line of sight to OB stars and velocities
of HI 21cm emission and absorption lines to estimate the peculiar velocity
distribution, after correction for solar motion and differential galactic
rotation.  These studies refer to fairly local gas, with distances less
than about 500-1000 pc. However with the advent and subsequent prominence
of molecular line observations, these studies were never repeated and were
in effect forgotten.  

         The results of these earlier studies are presented in Fig.1.  In
some cases the original histograms were only given in graphical form, and
these plots were converted to digitised images using a scanner, and then
measured on a computer terminal using standard image manipulation software.
 We estimate that the uncertainties in reproduction due to this procedure
are at the 10 percent level.  Since these studies contain a small number of
velocities compared to the present work, and because they are only meant to
be illustrative of the results already in the literature long ago, the
negative velocity portion of the pdf was reflected about v=0.

\cite{blaau52} (1952) was apparently the first to estimate the form of the
velocity pdf, using velocities of CaK lines from \cite{adams49}' (1949) catalog.
For 150 components along lines of sight to 120 stars with d$<$500 pc and 80
components toward 43 stars with d$>$500 pc, and excluding lines associated
with the Orion region, Blaauw showed that in both cases there were too many
components with high velocity to be fit by a single Gaussian, and that a
single exponential could better fit the data.  The probable existence of
blends due to the relatively poor velocity resolution of the Adams survey
was a problem, and Blaauw tried to correct his pdfs using an ingenious
technique that is
notable for its use of humans
as an analog random number generator.   The resulting pdf for the 150
nearby components
(Blaauw's Table 5a) is shown in Fig. 1 (filled circles) in log-linear form.
Notice that in
this form a Gaussian would be a parabola and an exponential would be a
straight line.
        \cite{takak67} (1967) also examined the K line velocities of Adams and
concluded that an exponential pdf gave a better fit than a Gaussian.  \cite{huang50}
(1950) and \cite{kapla54} (1954) preferred 1/v fits to the optical line pdf, based
again on Adams' catalogue.  (See Fig. 1 in \cite{kapla66} 1966).  \cite{siluk74}
(1974) examined the pdf of the
high-velocity ($v>20\ km\ s^{-1}$) tail of the Adams optical line sample
and fit the pdf with a power law of index
around --3.  \cite{munch57} (1957) presented new observations of optical
interstellar lines, at poorer resolution, toward
132 stars.  He used the ``doublet ratio method'' (CaH/K or NaD2/D1) to
compare curves of growth with calculations
based on Gaussian and exponential velocity pdfs.  The Gaussian fit required
a physically unrealistic increase of velocity dispersion with distance,
while the exponential provided a fit with a single mean speed, confirming
Blaauw's conclusion.

        Blending remained a problem with these interpretations.  Hobbs used
interferometric scans of NaD (\cite{hobbs69a} 1969a,b) and KI (\cite{hobbs74} 1974) 
in an attempt to fit individual profiles, and showed that most of Adams' lines
were probably multiple.  Hobbs found that the majority of individual line
profiles could be fit by Gaussians, while a smaller fraction favoured an
exponential, and many lines were not well-fit by either distribution;
however the velocity pdf was not examined.  \cite{falga73} (1973)
used Hobbs' NaD data to estimate the cloud peculiar velocity dispersion,
but the number of components was too small for an estimate of the pdf.
Later work on optical interstellar lines turned almost exclusively toward
studies of interstellar depletion patterns, and the question of the
velocity pdf was never re-examined.

\vspace{1in}\figcaption{Histogram representation of the radial velocity
probability density function (pdf) for distinct velocity components identified
in optical absorption line surveys along the line of sight to OB stars and HI
emission and absorption line surveys.  The data were either taken from tables
in the papers indicated, or measured from digitised scans
of graphs presented in the papers.  The negative-velocity portions of the pdfs
have been reflected about $v=0$.  In this log-linear plot a Gaussian pdf
would appear as a parabola while an exponential pdf would be a straight line.
Additional studies not shown in this figure are discussed in the text.}

        Meanwhile, HI 21cm survey data was accumulating.  \cite{takak67} (1967)
performed Gaussian decomposition and subtraction of solar motion and
galactic rotation, resulting in peculiar velocities for 544 HI emission
lines whose velocity pdf was presented for 3 intervals of linewidth and
different galactic latitude intervals.  He judiciously refrained from any
conclusions concerning the form of the pdf because of the dependence of the
derived velocity dispersions on galactic latitude.  (This was perhaps the
first paper to suggest consistency with the Kolmogorov velocity spectrum
for incompressible turbulence, on the basis of the power law scaling of
velocity dispersion with the sine of the latitude corresponding to a
scaling of the structure function with path length.)   The resulting
histogram for all 544 velocities is nevertheless shown in Fig. 1 (reflected
about v=0) as filled inverted triangles.  \cite{mast70} (1970)
presented a similar analysis for 268
peculiar radial velocities of HI 21 cm emission clouds that were
well-resolved in velocity.  Their Figure 4 clearly shows that an
exponential fits better than a Gaussian: a single Gaussian can fit the
wings or the core but not both. Their observed pdf is shown in Fig. 1 (open
circles).

        \cite{crovi78} (1978) studied the velocities of HI clouds identified as
components of Gaussian decomposition of absorption profiles observed
towards extragalactic sources in the Nancay survey.  About 300 such clouds
were selected on the basis of latitude, so that they might represent
relatively local gas.  The histogram of residual velocities, after
corrections for solar motion and differential galactic rotation, is given in
Crovisier's Fig. 1. The interpretation of the resulting histogram of
residual velocities is problematic because of the possibility that the
intermediate velocity features are spurious.  The histogram for all the
components is displayed in Fig. 1 (filled squares).  
\cite{dicke78} (1978) presented the residual
velocity distribution of clouds from the high-sensitivity and
high-resolution Arecibo 21 cm emission/absorption survey, but the number of
velocities is too small to estimate the functional form of the pdf.
Further discussion of the high-velocity tail of
the HI distribution and implications for turbulent energy requirements can
be found in \cite{kulka85} (1985) and \cite{lockm91} (1991).

All of the above studies included lines of sight that cover a
significant fraction of the sky and a range in distance, rather than
focusing on individual cloud complexes.
An exception is the detailed HI
emission mapping of two regions by \cite{versc74} (1974) using the NRAO 300 ft.
antenna.  Identifying ``clouds'' as distinct entities in spatial-radial
velocity space, Verschuur presented histograms of velocities for about 150
clouds in his region A and about 50 in region B.  The region B histograms
cannot be used to determine the intrinsic pdf of velocity fluctuations
without careful filtering (which we have not attempted) because of ordered
motion along the filamentary structures in the region.  The histogram for
region A, reflected about v=0, is shown as open squares in Fig. 1.

It should be emphasised that the rotation curve of the Galaxy and other
large-scale trends have been subtracted out in all of the studies mentioned 
above.  Furthermore, the results are likely not influenced by the 
Local Bubble, since the Verschuur sample was for a single region at a 
well-defined position and velocity, the optical-line clouds were detected 
toward OB stars which are on average about a kpc away, and the HI emission
and absorption studies likewise sample much larger distances (the Local 
Bubble only extends about 100 pc from the sun).  It is possible that
the optical line studies are influenced by gas motions in local 
HII regions around the OB stars, so some caution should be taken
in their interpretation.  However, it is not clear that the clouds
sampled in such studies lie in general at the same distance 
as the stars themselves (many studies have suggested otherwise), 
and the pdfs derived from the optical line data are similar 
to those derived from HI data, which are certainly not 
biased toward hot stars.

Overall, it can be seen from Fig. 1 that evidence for non-Gaussian, 
and probably exponential, centroid velocity pdfs in the relatively 
low-density atomic HI component of the ISM has existed for some 
time.  Many of these earlier works had already concluded that the
velocity pdf is better fit by an exponential (or 1/v) than a Gaussian.
It is surprising, and unfortunate, that all these early studies
have been in effect ``forgotten'' in more
current discussions of interstellar turbulence, since they suggest
that at least some forms of interstellar turbulence differ
significantly from incompressible turbulence, for which the velocity 
pdf is invariably nearly-Gaussian.  The implication is that the physical 
processes involved are substantially different than for incompressible 
turbulence.  This result for relatively low-density HI gas has recently 
received considerable support from the detailed  aperture synthesis 
HI study of the entire LMC by \cite{kim98} (1998).  For scales above 
the resolution limit of 15pc, the peculiar (fluctuation) velocity pdf 
of the gas (for motions out of the plane of the LMC) is extremely 
well-fit by an exponential, and a Gaussian seems to be 
certainly excluded.

We note that most of the HI studies reviewed in this section concern 
clouds which are probably not actively forming stars.  So there is good
evidence that, even in some ``quiescent'' regions, interstellar gas 
motions exhibit properties inconsistent with experiments and simulations 
of incompressible turbulence.  On the other hand, \cite{falga94} (1994) 
have found good agreement between the shapes (as quantified by moments of 
order 2 and 4) and point-to-point shape variations of observed 
spectral line profiles in quiescent molecular clouds and 
synthesised profiles from the ``nearly incompressible'' phase 
in simulations of mildly supersonic, decaying turbulence by
\cite{porte94} (1994).  
Apparently, either the physical processes occurring in at least some 
quiescent molecular clouds and HI clouds are substantially different, 
or the probability density function of centroid velocities contains 
different information than the second and fourth-order moments of 
individual line profiles.  The former posibility is likely, since the 
HI linewidths reported in the papers discussed above are significantly
supersonic in most cases.

We also point out that shapes and shape variations 
in optically thin line profiles similar those reported by
\cite{falga94} (1994) for mildly compressible turbulence simulations
have also been claimed for completely non-dynamical incompressible 
velocity fields (\cite{dubin95} 1995), highly supersonic simulations 
of cloud collisions (\cite{keto89} 1989), and highly supersonic 
stellar-driven self-gravitating turbulence fields by 
Vazquez-Semadeni et al. (Vazquez-Semadeni, private
communication).  For this reason we are not convinced that
the shapes and shape variations of individual line profiles
alone can be used to discriminate between models, even
for relatively quiescent clouds.  Additional diagnostics 
of the velocity field are therefore needed.

This section has been a review of centroid velocity pdfs 
observed in the diffuse atomic phase of the interstellar medium.  
In the remainder of this paper, we extend these studies to the 
higher-density molecular gas in active star-forming regions 
and also consider the pdf of velocity differences.  Ultimately 
our goal is to use these statistics of the velocity field to 
help understand the physical processes responsible for 
interstellar turbulence.

\section{Observations}\label{data}
The observations used in this paper are the same as those used in the 
previous statistical analyses of \cite{miesc94a} (1994, hereafter MB) 
and \cite{miesc95} (1995, hereafter MS), and include large-scale mappings 
of the giant molecular clouds in the northern part of the Orion region 
(Orion B) and in the Monoceros region, associated with the Mon R2 
infrared cluster.  Also included are observations of the smaller-scale 
clouds which appear in the Lynds catalog as numbers 1228 and 1551 
(\cite{lynds62} 1962), as well as observations of the molecular gas 
surrounding the Herbig-Haro object HH83.  In what follows, as in MS 
and in MB, we will refer to these five separate mappings as Orion B, 
Mon R2, L1228, L1551, and HH83.

\renewcommand{\floatpagefraction}{0.9}
\begin{deluxetable}{lcccccc}
\tablecaption{Observed Clouds\label{obs}}
\tablehead{\colhead{Region} & \colhead{R.A. (1950)} & 
\colhead{Decl. (1950)} & \colhead{Distance\tablenotemark{a} (pc)} 
& \colhead{Size (pc)} & \colhead{Mass\tablenotemark{a} ($\Msun$)} & 
\colhead{Mach Number\tablenotemark{b}}}
\startdata
Orion B & 05$^h$ 41$^m$ 08$^s$.1 
 & $-$01$^\circ$ 05$^\prime$ 00$^{\prime\prime}$ &
415 [1] & 20$\times$20\tablenotemark{c}, 45$\times$20 & 8$\times 10^4$ [6] 
 & 1.9--2.9\tablenotemark{c}, 4.0--7.1 \nl
Mon R2 &  06 05 22.0 & $-$06 22 25 & 830 [2] & 45$\times$40 & 9$\times
10^4$ [6] & 2.6--3.9 \nl
L1228 & 20 58 00.0 & $+$77 23 00 & 300 [3] & 2.6$\times$2.6 & 200--900 [7] & 2.6 \nl
L1551 & 04 28 40.0 & $+$18 01 42 & 140 [4] & 2.1$\times$1.5 & 80 [8] & 1.7 \nl
HH83 &  05 31 06.3 & $-$06 31 45 & 480 [5] & 0.9$\times$0.5 & 8--15 [9] & 1.5 \nl
\enddata
\tablenotetext{a}{References are as follows: [1] \cite{antho82} 1982,
[2] \cite{racin68} 1968, [3] \cite{greni89} 1989,
[4] \cite{elias78} 1978, [5] \cite{genze81} 1981,
[6] \cite{madda86} 1986, [7] \cite{bally95} 1995,
[8] \cite{snell81} 1981, [9] \cite{bally94} 1994}
\tablenotetext{b}{These Mach numbers are based on the centroid velocity fluctuation
  amplitudes listed in Table \ref{ReTa} ($\sigma_c$), and an assumed kinetic 
  temperature of 20K.}
\tablenotetext{c}{The first values listed correspond to regions 1a, 1b, and 1c, while the second
values represent the remainder: regions 2, 3, and 4 (see below).}
\end{deluxetable}
\begin{deluxetable}{lccccc}
\tablecaption{Regions Analysed\label{samsz}}
\tablehead{& \colhead{Number} & 
\colhead{Grid} & 
\colhead{Interpolation} & 
\colhead{Number} &
\colhead{LSR Velocity} \\
\colhead{Region} & \colhead{of} & 
\colhead{Spacing} & 
\colhead{Radius} & 
\colhead{of} &
\colhead{Integration Limits} \\
& \colhead{Spectra} & \colhead{(arcmin)} & 
\colhead{(arcmin)} & \colhead{Points} &
\colhead{(km s$^{-1}$)} }
\startdata
Orion B (1a) & 943 & 0.5 & 1.1 & 822 & [9.35,11.75] \nl
\phm{Orion B }(1b) & 1604 & 0.5 & 1.1 & 2094 & [7.5,10.5] \nl
\phm{Orion B }(1c) & 5817 & 0.5 & 1.1 & 4801 & [8.5,12.5] \nl
\phm{Orion B }(2) & 25510 & 0.5 & 1.1 & 18420 & [5.0,14.5], [9.2,11.9], \nl
& & & & & [7.0,12.1] \nl
\phm{Orion B }(3)\tablenotemark{a} & 23936 & 0.5 & 1.1 & 10433 & [3.0,9.2], [1.0,7.5] \nl
\phm{Orion B }(4) & 25311 & 0.5 & 1.1 & 21876 & [7.5,12.7] \nl
\tablevspace{.1in}
Mon R2 (1) & 3260 & 0.5 & 1.1 & 8630 & [10.6,13.4] \nl
\phm{Mon R2 }(2) & 2813 & 0.5 & 1.1 & 7592 & [7.7,12.5] \nl
\phm{Mon R2 }(3) & 4570 & 0.5 & 1.1 & 12208 & [11.2, 14.4] \nl
\tablevspace{.1in}
L1228 & 1922 & 0.5 & 1.1 & 3532 & [5.5,8.1] \nl
L1551 & 2327 & 0.5 & 1.1 & 4726 & [-9,-6] \nl
HH83 & 1142 & 0.05 & 0.25 & 6235 & [5.0,7.5]
\enddata
\tablenotetext{a}{As discussed in the text, region 3 in Orion B
overlaps spatially with regions 2 and 4 and therefore
includes most of the same
spectra, but integrated over a different velocity range.}
\end{deluxetable}

The Orion B, Mon R2, L1228, and L1551 observations are
emission-line measurements of the 110 GHz, $J = 1\rightarrow0$
transition in $^{13}$CO.  They were obtained with the AT\&T Bell
Laboratories 7 m offset Cassegrain antenna located in Holmdel,
New Jersey, between 1985 December and 1991 June.  This
antenna has a very clean Gaussian beam with a FWHM of
100$^{\prime\prime}$ between 98 and 115 GHz and the effective
channel bandwidth (resolution) is 100 kHz (0.27 km s$^{-1}$)
for the Orion B, Mon R2, and L1228 data sets and 50 kHz
(0.14 km s$^{-1}$) for the L1551 data set.  Typical
signal-to-noise ratios are between 5 and 9.  Further details
on the observations and measurement techniques can be found
in \cite{bally87} (1987), \cite{bally87b} (1987), 
\cite{bally89} (1989), \cite{pound90} (1990),
\cite{bally91} (1991), \cite{pound91} (1991), 
\cite{bally95} (1995), and MB.

The observations of the HH83 molecular cloud were obtained with the
IRAM-30m telescope on Pico Veleta, Spain in April and December 1989,
and have a higher spatial resolution (half power beamwidth
= 13$^{\prime\prime}$) and smaller field of view.
The mapped region lies just west of the Orion A molecular cloud
(L1641) and is associated with the Herbig-Haro object HH83.
The measurements are of the $^{13}$CO $J = 2\rightarrow1$ emission
line at 220 GHz, with an effective frequency resolution of 100 kHz
(0.13 km s$^{-1}$).  A typical signal-to-noise ratio for
these observations is about 20.  More information on the
data acquisition, as well as the structure, environment, and
physical conditions in the molecular cloud can be found in
\cite{bally94} (1994).

	We recognise that $^{13}$CO is an imperfect tracer of the gas
distribution and velocity field in dense molecular clouds.  The primary
problem is the likelihood that the $^{13}$CO transitions are optically thick
along lines of sight with the largest total column density.  Studies of the
correlation of $^{13}$CO line strength with visual and infrared extinction and
with rarer CO species such as C$^{18}$O (\cite{frerk82} 1982;
\cite{frerk89} 1989; \cite{lada94} 1994) suggest that $^{13}$CO saturates at
visual extinctions greater than about 5 mag, although these papers only
study a small number of individual clouds.  Furthermore, the comparison of
C$^{18}$O emission with infrared extinction by \cite{alves98} (1998)
suggests that for A$_v \gtrsim$10 mag, C$^{18}$O underestimates the total optical depth,
possibly due to depletion of CO onto grain surfaces.  These results imply
that the present observations are biased in the sense that they do not
penetrate the highest-column density substructure.  However, the fractional area 
covered by column densities greater than A$_v \approx$ 5 is small, about
five or six percent in Orion B, one or two percent in Mon R2, and 
effectively zero in L1228, L1551, and HH83 (assuming the following 
conversion factors: 
I($^{13}$CO)$\rightarrow$N($^{13}$CO) = 5.9$\times 10^{14}$ cm$^{-2}$ K$^{-1}$ km$^{-1}$ s,
N($^{13}$CO)$\rightarrow$N(H$_2$) = 7$\times 10^{5}$, and
N(H$_2$)$\rightarrow$A$_v$ = 1.06$\times 10^{-21}$ mag cm$^{2}$).  So, we are probably probing 
{\em most} of the molecular gas in the surveyed regions.  Furthermore, it 
would be extremely time-consuming to collect, say, C$^{18}$O spectra for 
the large number of lines of sight (more than 75,000) examined here, because 
of the required sensitivity.  So, in terms of spatial coverage and statistical 
characterisation of the velocity field, $^{13}$CO is probably the best probe 
available.  Nevertheless, the fact that this tracer probably does not probe 
the highest-column density substructure should be
kept in mind in interpreting the statistical results of \S4 below.

\begin{figure*}[t]
\vspace{1in}\figcaption{Shown are the Orion B $^{13}$CO $J = 1\rightarrow0$
observations used in the analysis to follow.  Contours represent
integrated intensity and images represent centroid velocity.
Red and blue in the centroid velocity images correspond respectively
to receding and approaching motions relative to the mean LSR velocity
of the cloud (see the color table in Fig. \ref{vimM}).  
The data set has been subdivided into 6 distinct
regions as described in the text.  The orientation of the coordinate
system with respect to the north and west directions on the
plane of the sky is indicated.  All coordinates represent the offset
in arcminutes from the position of the origin, which is listed in
Table \ref{obs} and applies to all regions.  The dashed blue line
indicates the position of region 3, which overlaps regions 2 and 4.
These, together with the dotted white lines, also serve to delimit 
the composite portions of regions 2 and 3 which were obtained using
different integration limits as described in the text. 
The LSR velocity integration limits used for each region are 
indicated in brackets\label{vimO}}
\vspace{1in}\figcaption{Similar to Figure \ref{vimO},
but for the $^{13}$CO $J = 1\rightarrow0$ Mon R2, L1228, and L1551
observations and the $^{13}$CO $J = 2\rightarrow1$ HH83 observations.
The color table used for the centroid velocity images (which applies 
also to Fig. \ref{vimO}), is displayed on the right.
The orientation of the coordinate system is indicated
in each panel and all axes represent arcminute offsets with respect
to the positions listed in Table \ref{obs}.\label{vimM}}
\end{figure*}

The observations are presented in Figures \ref{vimO} and \ref{vimM}.
The coordinates corresponding to the origin (0,0) of each map are listed in
Table \ref{obs} (see also MB, Table 1).  The centroid
velocity\footnote{The centroid velocity in each line of sight is defined
as $\Sigma T_i v_i / \Sigma T_i$, where $T_i$ and $v_i$ are the brightness
temperature and velocity corresponding to the $i^{th}$ spectrometer
channel, and the summations span some selected velocity range.}
along each line of sight is indicated by the color image,
with blue and red denoting movement respectively toward and away
from the observer with respect to the mean LSR velocity of the cloud.
The contour plot overlays indicate the integrated intensity in
the $^{13}$CO transition considered (see above).

The number of independent spectra included in these observations totals 
more than 75,000 (see Table \ref{samsz}).  The spectra for each region
were extracted, reduced, and interpolated onto a regular spatial grid
as described by MB.  The grid spacing and interpolation radius
used for each set of observations is listed in Table \ref{samsz}.
The velocity integration ranges for each region, also listed
in Table \ref{samsz}, were chosen to be large enough to span most 
of the emission, but still small enough to mitigate the influence 
of instrumental noise on the centroid fluctuations.  In the case of 
Orion B and Mon R2, some of the integration ranges were further 
restricted in order to isolate particular cloud components.
See MB for further details.

After interpolating the spectra onto regular spatial grids, threshold 
integrated intensity levels were introduced in order to eliminate pixels 
with low signal-to-noise.  Any remaining low-intensity emission located 
outside the main cloud boundaries and having dramatically different 
centroid velocity values (likely influenced by noise) was also removed.  
The number of pixels in each of the resulting centroid velocity maps in 
Figures \ref{vimO} and \ref{vimM} are listed in Table \ref{samsz}.
Since the maps were oversampled, the number of points in the centroid
and integrated intensity maps is in many cases larger than the 
associated number of independent spectra.

Several distinct cloud components, or ``regions'', distinguished by
their spatial position and their mean LSR velocity, emerge when the
Orion B and Mon R2 data cubes are analysed closely.  In the study presented
here, as in MS and in MB, we have divided these data
sets accordingly and have considered each of the components separately.
Thus, in Figures \ref{vimO} and \ref{vimM}a we have indicated six regions
(1a, 1b, 1c, 2, 3, and 4) in Orion B and 3 regions (1, 2, and 3) in Mon R2.
In order to facilitate the isolation of specific cloud components in the
Orion B and L1551 observations, we have chosen coordinate systems which
are tilted somewhat with respect to the right ascension - declination
standard.  The orientation of each map is indicated by arrows in
Figures \ref{vimO} and \ref{vimM}.

\begin{figure*}[b]
\vspace{1in}\figcaption{Shown are grey scale images of the centroid
velocity fluctuations in Orion B, which were obtained by filtering
the raw centroid velocity maps shown in Figure \ref{vimO} as
described in the text.    Increasing brightness
corresponds to a velocity directed increasingly away from the
observer, as indicated by the intensity table in Fig.
\ref{vcM}.\label{vcO}}
\vspace{1in}\figcaption{Similar to Figure \ref{vcO}, but for the
Mon R2, L1228, L1551, and HH83 data sets.  These maps were obtained by 
applying spatial filters to the centroid velocity maps shown in 
Figure \ref{vimM} (see text).  The sense of the grey scale is 
shown by the intensity bar, which also applies to Figure \ref{vcO}.\label{vcM}}
\end{figure*}

All the coordinate axes in Figure \ref{vimO} are with respect to the same
origin in order to facilitate the placement of the different regions
for the reader.  Thus, region 1 (a, b, and c), lies directly above region 2
in the coordinate system chosen (see MB, Fig. 1e),
or in other words, just northeast of the main Orion B cloud.  It has
been displaced and enlarged somewhat in Figure \ref{vimO} for clarity.
Region 3 overlaps regions 2 and 4 in the manner indicated by the
dashed blue lines.   It appears to be a fairly
well-defined separate cloud component which lies at a significantly lower
LSR velocity (by $\sim$ 4 km s$^{-1}$) than the remaining emission and
was isolated by choosing appropriate velocity integration limits
(see below and MB).  It is likely associated with the doubly-peaked
$^{12}$CO line profiles in this region reported by \cite{madda86}
(1986).  In order to isolate the spatially overlapping regions in Orion B,
we further subdivided regions 2 and 3 into portions, each integrated
over a somewhat different velocity range, as indicated in Figure
\ref{vimO}.  These portions were then combined to produce composite
maps for regions 2 and 3.
  
The cloud components used for Mon R2 are shown in the three separate
panels of Figure \ref{vimM}a.  As in the case of Orion B, they 
were chosen on the basis of both spatial connectivity and mean 
LSR velocity using channel maps and spatial-velocity diagrams.
The relative orientation of these regions can be discerned
by noting the coordinate axes in each panel (which are
all with respect to the same origin) and by referring to
Figure 1d of MB.  The distinction between the three different 
regions defined here is also apparent in the $^{12}$CO channel
maps presented by \cite{xie94} (1994; their Figures 1 and 2), who
interpret the relative blueshift of the southeastern portion of the
cloud (our region 1) as evidence for a large-scale, expanding shell 
(see \S\ref{env_mon} in Appendix \ref{overview}).

An overview of the physical conditions and environment in each of the
molecular clouds in our study is given in Appendix \ref{overview}.
Briefly, our sample covers a broad range of scales and conditions 
relevant to the velocity field, with energy input ranging from hot massive 
stars to low-mass protostellar outflow sources.  What the regions have 
in common is that they all contain strong, active sources of momentum 
and energy to drive the ``turbulence'' whose statistics we wish to study.

For reference, Table \ref{obs} lists estimates for the distance to each
region, and for the total mass and plane-of-the-sky extent
of the CO-emitting gas.  Also listed are estimates for the Mach 
number of each region, based on the measured rms centroid velocity 
fluctuation (see Table \ref{ReTa} below) and on an assumed
sound speed of 0.287 \kms (corresponding to a kinetic 
temperature of 20K).  Note that the range is wide, from
1.5 to 7.1, and that all regions are substantially supersonic,
suggesting that compressibility may play an important role
in their dynamics.

\section{Statistical Analysis}\label{results}
\subsection{Centroid Velocity Fluctuations}\label{vc}
\subsubsection{Analysis}\label{analysis}
It is apparent in the centroid velocity images shown in
Figures \ref{vimO} and \ref{vimM} that many of the regions
studied have large-scale velocity gradients across them,
which at least in some cases, dominate any statistical
quantifiers which may be computed.  Since statistical
approaches generally lose spatial information by
considering only distributions or averages, their
utility depends on some degree of uniformity in
the data.  Any statistical quantifier which is
dominated by a small number of large-scale features
in the data can give misleading results.  In other
words, large-scale gradients should be removed from
the centroid data in order for our statistical results to 
be meaningful and comparable to theoretical models, numerical 
simulations, and laboratory experiments involving 
turbulent fluids.

\begin{figure*}[b]
\vspace{1in}\figcaption{Centroid velocity pdfs for Orion B, corresponding to the images
shown in Figure \ref{vcO}.  Crosses denote the histogram estimate with
statistical ($N^{1/2}$) error bars and solid and dashed lines represent
the Johnson and adaptive kernel estimators discussed in the text.
The corresponding Johnson parameters are listed in Table \ref{John_vc}.
\label{vc_pdfsO} }
\vspace{1in}\figcaption{Similar to Figure \ref{vc_pdfsO}, but for the Mon R2, L1228, L1551,
and HH83 data sets shown in Figure \ref{vcM}.  The parameters corresponding
to the Johnson estimators (solid lines) are listed in Table \ref{John_vc}.
\label{vc_pdfsM}}
\end{figure*}

For these reasons, we have removed large-scale gradients
by first applying a low-pass filter, or smoothing function,
to each data set and then subtracting this smoothed map from
the original to obtain the centroid velocity fluctuations.
The filters were chosen to be as wide as possible,
while still eliminating any prominent, extensive, antisymmetric
``lobes'' in the autocorrelation function of the fluctuation
maps.   Such features in the autocorrelation function are
the characteristic signature of large-scale gradients in
the data.  The details of the filtering process are described
by MB and will not be discussed further here.  

All of the analysis presented in this and the following sections
(\S\ref{vc}--\S\ref{Resec}) is based on the centroid velocity
fluctuations, which are shown as Grey-scale 
images in Figures \ref{vcO} and \ref{vcM}.  Although the pdfs of 
the centroid velocity fluctuations (discussed below) generally 
appear significantly different from the unfiltered data, they are 
not sensitive to the precise value of the effective cutoff wavelength 
or the shape of the filter.  Furthermore, since the analysis of 
velocity differences in \S\ref{dvc} primarily samples small-scale 
structure, it is very insensitive to spatial filtering.  

Notice that although some of the highest-amplitude fluctuations, which 
will appear in the far tails of the corresponding pdfs, are found within 
the main bodies of the mapped regions, others occur near the cloud
edges or in ``outlier'' regions--areas which could not be observed 
over large continuous spatial extents due to the sensitivity limits 
of the observations or the influence of overlapping cloud components.
For example, this behaviour can be seen in the L1551 and Orion B, 
region 1b maps in Figures \ref{vcO}c and \ref{vcM}e.  Such high-amplitude
edge or outlier points introduce a bias in the far tails of the
estimated pdfs if 
they represent emission that is noise-dominated or kinematically 
distinct from the main body of each region.  Although 
much of this emission was removed from the maps prior to analysis, 
it is very difficult to identify and remove all of it.  The centroid 
velocity pdf estimates therefore still contain some of this bias.  
The influence of noise and overlapping cloud components is discussed 
further in Appendix \ref{noise}.

The pdfs, or probability density functions, corresponding to each of the
centroid velocity fluctuation images of Figures \ref{vcO} and \ref{vcM} are 
shown in Figures \ref{vc_pdfsO} and \ref{vc_pdfsM}.  Although the classical 
histogram is a familiar and robust estimator of the pdf, it is generally 
not the most accurate.
\cite{vio94} (1994) have investigated several alternative pdf
estimators in addition to the histogram and have compared their
performance on some typical astronomical problems.  They conclude
that the histogram is decidedly inferior in the applications they
considered and recommend the use of more sophisticated pdf
estimators.  Motivated by this result, we have implemented two of
the histogram alternatives described by \cite{vio94} (1994): the
adaptive kernel estimator and the Johnson estimator.

Kernel estimators approximate the pdf at each data value in terms
of a sum over all data points, weighing each according to a
specified window function, or kernel.  They do not require
binning of the data as in the classical histogram estimator
and are generally less sensitive to noise in the data,
producing a smoother, continuous, and often more reliable result.
In the adaptive kernel method, the effective width of the
kernel increases systematically where the data density is low,
thereby improving accuracy in the pdf tails.  In our implementation,
the form of the kernel and the manner in which its width is allowed
to vary are as described by \cite{vio94} (1994).  The adaptive kernel
estimation of each of the pdfs in Figures \ref{vcO} and \ref{vcM}
is shown as a dashed line.

An alternative strategy for pdf estimation is based on defining some general
class of analytic functions involving several parameters which can be
approximated using statistical moments or percentiles derived from the data
set.  The advantage of such an approach is that it does not require the 
subjective binning of data or the choice of a kernel function and window 
width. It can also be
less sensitive to noise in the data than either histogram or kernel methods.
The disadvantage of parametric methods is that, however general, they
impose some functional form to the data which may not be present.
Our implementation of the Johnson system follows that described
by \cite{vio94} (1994).  Since the centroid velocity is in principle
an unbounded variable, we consider only the so-called $S_U$ family of Johnson
estimators, which is characterised by the four parameters
$\eta$, $\epsilon$, $\lambda$, and $\gamma$, and which is defined by
the following functional form:
\be{JSU}
f_{J}(v) =
\frac{\eta}{\sqrt{2\pi\left[(v-\epsilon)^2+\lambda^2\right]}} \; \times
\hspace{1.3in}
\ee
\begin{displaymath}
\hspace{.25in}
\exp \left\{- \frac{1}{2}\left[\gamma + \eta \ln
\left(\frac{(v-\epsilon)}{\lambda}
+ \sqrt{\left(\frac{v-\epsilon}{\lambda}\right)^2+1}\right)\right]^2 \right\}
\;
\end{displaymath}
for $-\infty < x < \infty$.
The Johnson estimation for each of the pdfs in Figures \ref{vcO} and
\ref{vcM} is shown as a solid line, and the best-fit parameters are
listed in Table \ref{John_vc}.

\begin{deluxetable}{lcccc}
\tablecaption{Johnson Parameters for the Centroid Velocity
pdfs\label{John_vc}}
\tablehead{\colhead{Region} & \colhead{$\eta$} & 
\colhead{$\epsilon$} & \colhead{$\lambda$} & 
\colhead{$\gamma$} }
\startdata
Orion B (1a) & 1.55 & -1.63$\times 10^{-2}$ & 0.192 & -0.114 \nl
\phm{Orion B }(1b) & 5.20 & -0.292 & 1.01 & -0.292 \nl
\phm{Orion B }(1c) & 1.17 & 0.218 & 0.285 & 0.504 \nl
\phm{Orion B }(2) & 2.03 & 0.149 & 0.534 & 0.670 \nl
\phm{Orion B }(3) & 2.08 & -0.318 & 1.13 & -0.604 \nl
\phm{Orion B }(4) & 1.32 & -7.75$\times 10^{-2}$ & 0.476 & -0.157 \nl
\tablevspace{.1in}
Mon R2 (1) & 5.37 & 1.26 & 1.46 & 4.31 \nl
\phm{Mon R2 }(2) & 4.27 & 0.121 & 0.929 & 0.609 \nl
\phm{Mon R2 }(3) & 2.80 & -6.70$\times 10^{-2}$ & 0.769 & -0.349 \nl
\tablevspace{.1in}
L1228 & 1.64 &  5.00$\times 10^{-2}$ & 0.175 & 0.322 \nl
L1551 & 1.11 & -5.65$\times 10^{-2}$ & 0.141 & -0.201 \nl
HH83 & 1.59 & 5.29$\times 10^{-4}$ & 0.158 & 0.255
\enddata
\end{deluxetable}

The centroid velocity pdfs presented by MS
apply to the same data as the pdfs in Figures \ref{vc_pdfsO} and 
\ref{vc_pdfsM},
but they were based solely on the histogram pdf estimator (with a
slightly different binning than that used here).  
MS also constructed several composite pdfs, one of which was formed from
the six subregions of Orion B, another from the three subregions of
Mon R2, and a third composite pdf which was composed of all twelve
data sets.  For the sake of brevity, we have chosen to omit these
composite pdfs from the present study, although it is worth
noting that the Orion composite pdf in particular appears very
nearly exponential.

Most of the pdfs in Figures \ref{vc_pdfsO} and
\ref{vc_pdfsM} exhibit significantly non-Gaussian shapes,
a result which will be discussed in \S\ref{vcdisc} below.
However, before proceeding, we note that a few pdf features
probably originate from the superposition of two or more
distinct cloud components.  For example, the morphology and
kinematics of regions 1b and 1c in Orion B (see Figs.
\ref{vimO} and \ref{vcO}) suggest several cloud components,
which may be the source of some of the bimodality
and fine structure exhibited by the associated pdfs
(Figs. \ref{vc_pdfsO}b and \ref{vc_pdfsO}c).  Although 
the spatially overlapping cloud components in Orion B regions 
2 and 3 were separated out by choosing appropriate velocity 
integration limits, some contamination remains.  Thus, the 
low velocity ($< -2$ \kms) pdf tail in region 2 
(Fig. \ref{vc_pdfsO}d) and the associated large kurtosis value of 11 
(see Table \ref{moms_vc} below), are probably influenced by residual 
emission from region 3, and the abrupt high-velocity ($\sim$ 2 \kms) 
cutoff in the region 3 pdf (Fig. \ref{vc_pdfsO}e) is likely 
influenced by the imposed upper integration limit.
Also, residual overlap in Mon R2 may account for some of the 
excess in the high-velocity tail of region 1 and the low-velocity 
tail of region 2 (Figs. \ref{vc_pdfsM}a and b).

Another way to quantify the spread and shape of probability density
functions is by the sample moments, computed directly from the data
set.  We consider the first four central moments, defined as follows:
\be{M1}
\mbox{mean} = \mu = \frac{1}{N} \Sigma_{x, y} v(x,y) \; ;
\ee
\be{M2}
\mbox{standard deviation} = \sigma = \sqrt{\frac{1}{N} \Sigma_{x, y}
\left[v(x,y)
- \mu \right]^2} \; ;
\ee
\be{M3}
\mbox{skewness} = \frac{1}{\sigma^3}\frac{1}{N} \Sigma_{x, y}
\left[v(x,y) - \mu \right]^3 \; ;
\ee
\be{M4}
\mbox{kurtosis} = \frac{1}{\sigma^4}\frac{1}{N} \Sigma_{x, y}
\left[v(x,y) - \mu
\right]^4 \; .
\ee
The summation over $x,y$ spans the area of the cloud or in other words,
the entire data set.
The first two moments quantify the location and spread of the pdf, while the
third and fourth are nondimensional quantities which contain information
about its shape.  In particular, the skewness and kurtosis are measures of the
symmetry and ``flatness'' of the pdf respectively.  A Gaussian distribution
has a kurtosis of 3, and a value larger than 3 implies that the
pdf in question has relatively more prominent tails -- in other words,
high-amplitude events are more numerous than would be expected for a
Gaussian random variable.  In the turbulence literature this
behaviour is usually referred to as
``intermittency'' (although the same term is often used in reference to the
variation of the scaling exponents of the
moments of the velocity {\it difference} distribution).
 A kurtosis less than three implies the
opposite--that pdf tails are less prominent relative to a Gaussian
distribution.
An exponential distribution ($f(x) \propto \exp (-\vert x \vert)$) has a
kurtosis equal to 6.  Table \ref{moms_vc} lists the second,
third,
and fourth moments for each of the pdfs shown in Figures \ref{vc_pdfsO} and
\ref{vc_pdfsM}.  Note that the skewness and kurtosis values listed are
central sample moments, and are therefore slightly different from the values
given by MS, which were computed using the histogram as a
pdf estimator.

\begin{deluxetable}{lcccc}
\tablecaption{Moments and Stretching Exponents for the Centroid Velocity pdfs\label{moms_vc}}
\tablehead{\colhead{Region} & \colhead{Standard Deviation} & 
\colhead{Skewness} & \colhead{Kurtosis} & \colhead{$\beta$} \\
& \colhead{(\kms)} }
\startdata
      Orion B (1a) & 0.15 & -0.018 & 2.9 & 2.1 \nl 
\phm{Orion B }(1b) & 0.22 & -0.50 & 4.8 & 1.2 \nl 
\phm{Orion B }(1c) & 0.36 & -0.58 & 4.2 & 1.4 \nl 
\phm{Orion B }(2)  & 0.34 & -1.3 & 11 & 0.71 \nl 
\phm{Orion B }(3)  & 0.63 & 0.073 & 4.0 & 1.4 \nl 
\phm{Orion B }(4)  & 0.46 & 0.12 & 4.9 & 1.2 \nl 
\tablevspace{.1in}
       Mon R2 (1)  & 0.37 & -0.18 & 3.0 & 2.0 \nl 
\phm{Mon R2 }(2)  & 0.23 & -0.23 & 3.2 & 1.8 \nl 
\phm{Mon R2 }(3)  & 0.29 & 0.14 & 3.3 & 1.8 \nl 
\tablevspace{.1in}
   L1228           & 0.14 & -0.17 & 6.0 & 1.0 \nl 
   L1551           & 0.18 & 0.60 & 4.6 & 1.2 \nl 
   HH83            & 0.12 & -0.39 & 4.2 & 1.3 \nl 
\enddata
\end{deluxetable}

Many theoretical and empirical studies involving pdfs in the context of 
incompressible turbulence are concerned with stretched exponential forms, 
whereby the pdf tails fall off as 
$f(x) \propto \exp\left\{ - a \vert x \vert^\beta \right\}$ 
(e.g. \cite{kaila92} 1992; \cite{lohse93} 1993; 
\cite{yee97} 1997; \cite{frisc97} 1997).  
A Gaussian pdf is characterised by $\beta = 2$, and an
exponential pdf by $\beta = 1$.  Fractional $\beta$ values can arise from 
random multiplicative processes and represent a straightforward and
useful way of quantifying the departure of a pdf from Gaussian
statistics.

The value of the stretching exponent $\beta$ determines how quickly 
the pdf decays, and as a result, how prominent the pdf tails are.  
For a properly normalised stretched exponential pdf, 
there is a one-to-one correspondence between the value of $\beta$ and the 
pdf kurtosis.  Therefore, if the observed velocity centroid pdfs can be 
approximately described by stretched exponential forms, then the sample kurtosis 
values listed in Table \ref{moms_vc} imply an effective $\beta$ for each data set.
The results, listed in Table \ref{moms_vc}, yield stretching exponents
for Orion B and the smaller clouds (L1228, L1551, and HH83) which 
generally range between 1.0 and 1.4.  Exceptions include Orion B
regions 1a ($\beta=2.1$) and 2 ($\beta=0.71$).  The Mon R2 regions
yield higher $\beta$ values, between 1.8 and 2.  Curve fits to 
the tails of the Johnson pdf estimators yield similar, although 
slightly lower results for the stretching exponents.  Adopting 
a Poisson weighing and combining the high and low velocity 
tails to improve statistics, such curve fits, excluding 
Mon R2, yield $\beta$ values ranging from 0.77 to 1.2, with 
the exception of Orion B, region 1b ($\beta=1.6$).  The results 
for Mon R2 again indicate somewhat steeper tails, 
with 1.5 $< \beta < $ 1.7.

\begin{figure*}[t]
\vspace{1in}\figcaption{The tails of the centroid velocity pdfs in Figures \ref{vc_pdfsO} 
and \ref{vc_pdfsM} are here shown on log-log axes in order to emphasise
any power law components ($\propto v^{n}$) which may be present.  
On such axes, power law forms would appear as straight lines.  
The abscissa in each panel runs from one to six times the pdf standard 
deviation, $\sigma$ (see Table \ref{moms_vc}).  Solid lines and asterisks 
denote the high velocity pdf tails and dashed lines and squares denote 
the low velocity pdf tails.  All pdf tails are normalised to unity 
at a velocity of one $\sigma$.  Plot symbols (asterisks and squares) 
represent the histogram pdf estimator, while lines (solid and dashed) 
represent the adaptive kernel pdf estimator.  The dotted line in the lower 
left corner of each panel illustrates a power law with a similar
slope for comparison.\label{loglog}}
\end{figure*}

We have also considered the possibility that power law forms,
$f(x) \propto x^n$, may fit at least some of the pdf tails better 
than stretched exponentials.  Power law velocity distributions could 
arise from stellar winds, outflows or superbubbles 
(e.g. \cite{silk95} 1995; \cite{oey98} 1998) and would suggest
physical processes substantially different than those occurring 
in isotropic, incompressible turbulence.  When the pdf tails 
of Figures \ref{vc_pdfsO} and \ref{vc_pdfsM} are plotted
on log-log axes rather than log-linear axes (Fig. \ref{loglog}), 
many of them appear linear in portions, suggesting power laws.  
This is particularly the case for L1228, HH83 and 
Orion B, region 4, and possibly also in L1551 and 
Orion B, regions 1a, 2, and 3.  Curve fits to the pdf estimators in these regions 
show some evidence for power law indices of $\approx -4.5 \pm 1$, although 
the positive velocity tails in L1551 and Orion B, region 3 seems significantly
shallower ($n \approx 2$ or $3$).  Power law forms are displayed
in each of the frames of Fig. \ref{loglog} for comparison.  
Although stretched exponentials in most (but not all; c.f. L1228) cases 
appear to provide somewhat better fits to the observed pdf tails than power laws, 
the data are generally not of high enough quality to make a reliable distinction.

\subsubsection{Discussion}\label{vcdisc}
In most cases, there is good agreement between the three different pdf
estimators (histogram, adaptive kernel, and Johnson) shown in Figures
\ref{vc_pdfsO} and \ref{vc_pdfsM}.  Exceptions include the pdf excesses
in the tails of Orion B regions 1b and 2, and the fine structure in Orion B,
region 1c, all of which appear in the histogram and adaptive kernel
estimators, but are not well-represented by the Johnson system.   
In general, however, the Johnson system provides a reasonable analytic 
approximation to the pdfs presented here, and the parameters listed in 
Table \ref{John_vc}, together with the analytic form given by 
equation (\ref{JSU}) can therefore be used by observers and modellers 
alike to roughly reproduce and compare with our results.

The velocity pdfs for the Mon R2
subregions and the high-velocity portion of Orion B, region 1b
appear nearly parabolic on the log-linear axes shown, implying Gaussian
forms, although the far tails for these regions still may suggest
exponentials or power laws.  However, most of the other pdfs appear distinctly 
non-Gaussian.  Several of the pdf tails, particularly for regions 2, 3, and 4 
in Orion B, as well as L1228, L1551, and HH83, instead suggest nearly 
exponential (straight lines on the log-linear axes of \ref{vcO} and \ref{vcM}), 
or possibly power-law (straight lines on the log-log axes of Figure \ref{loglog}) 
forms.  These results are
consistent with the derived stretching exponents, which lie near 2 for 
the three Mon R2 regions, and near 1.1 $\pm 0.3$ for most of the 
others (see Table \ref{moms_vc} and the associated discussion).
In no case do we find pdf tails which decay more rapidly than 
Gaussian ($\beta \lesssim 2$ for all regions).

It should be noted that the more Gaussian nature of the Mon R2 
centroid velocity pdfs is not associated with particularly high 
or particularly low Mach numbers relative to the nine other
regions (Table \ref{obs}). However, relative to regions 2, 3, 
and 4 in Orion B, which have a comparable size and mass scale, the 
Mon R2 regions do possess somewhat lower Mach numbers (Table \ref{obs}) 
and somewhat higher Reynolds numbers (Table \ref{ReTa}).
Also, differences in the distribution and dynamical influence 
of young, massive stars in Orion B and Mon R2 may be significant
(see Appendix \ref{overview}).

The pdf shapes can be described further using the kurtosis values listed
in Table \ref{moms_vc}.  The three Mon R2 regions are all characterised
by a kurtosis of about 3, again suggesting nearly Gaussian statistics.
However, with the exception of Orion B, region 1a, the other pdfs have 
larger kurtosis values, ranging from 4 to 6 (As explained above, the
large kurtosis of 11 found for Orion B, region 2, is probably anomalous, 
due to overlapping with region 3).
These results are in strong contrast  to those for incompressible turbulence,
where the velocity pdf is nearly  Gaussian (see references given in
\S\ref{background}), and they imply  a fundamental difference between the
physics of interstellar supersonic,
driven turbulence and incompressible turbulence.  An interpretation of the
difference based on the fact that kinetic energy is not a conserved
quantity in supersonic turbulence is discussed in Paper III.

Our results for molecular gas powered by internal sources are basically in
agreement with the velocity pdfs from
optical line and HI emission and absorption components for cool atomic gas
displayed in Fig. 1.  Since these
latter regions are generally not self-gravitating and contain no internal
power sources, the similarity may imply that the non-Gaussian behaviour is
either a result of nonlinear advection or that the latter regions possess
turbulence that is powered by external sources.

However, we note that at least 3 of the 11 subregions studied here
{\em do} appear Gaussian, so our results should not be taken as an 
indication that all molecular clouds have exponential or even non-Gaussian 
velocity distributions.  However since the 3 regions in question are all 
in the same GMC (Mon R2), we can say that the Gaussian distributions 
are definitely in the minority.  A similar statement can be made for the
optical and HI line results discussed in sec.2.2.  Perhaps the most
interesting challenge, theoretically, is to understand why the pdf is
approximately exponential in most of the regions but apparently 
Gaussian in Mon R2.

\subsection{Centroid Velocity Differences}\label{dvc}
As mentioned in \S\ref{background}, probability density functions
of velocity derivatives or differences have proven to be an
important and well-studied flow diagnostic in the context
of incompressible turbulence, where they exhibit significantly
non-Gaussian shapes.  It is therefore of great interest
to construct maps of centroid velocity differences in molecular
clouds and investigate their statistical properties.
In this section we apply such an analysis to our data sets.

\begin{figure*}[t]
\vspace{1in}\figcaption{Shown are grey-scale images of the centroid velocity
differences for each subregion in Orion B, corresponding to a lag in pixels
of $\ttau = [1,1]$.
As indicated by the legend in Figure \ref{dvM}, the grey scale has only three
shades, chosen to highlight the spatial distribution of large-amplitude
difference values (white).\label{dvO}}
\vspace{1in}\figcaption{Similar to Figure \ref{dvO}, but for the
Mon R2, L1228, L1551, and HH83 data sets.  Again, all maps correspond
to a lag in pixels of $\ttau=[1,1]$.  The grey-scale intensity levels
are based on the standard deviation of each map, as indicated by the
legend.\label{dvM}}
\vspace{1in}\figcaption{Shown are grey-scale images of the centroid velocity
differences for Orion B, region 4 at the lags indicated (measured 
in pixels).  The dashed boxes indicate the displacement as 
specified by the lag vector, $\ttau$.  The sense of the grey-scale 
is the same as in Figures \ref{dvO} and \ref{dvM}.\label{dvO4}}
\end{figure*}

The velocity differences we consider are defined as follows;
\be{DeltaV}
\Delta v ({\bf x}, \ttau) = v({\bf x}) - v({\bf x} + \ttau) \; ,
\ee
where ${\bf x}$ and $\ttau$ are two-element vectors corresponding to
position on the plane of the sky and $v$ is the centroid velocity.
The displacement $\ttau$ is referred to as the two-dimensional
vector lag.  The structure function is just the mean square of
these velocity differences, averaged over the area of the cloud
and normalised with respect to the centroid velocity variance
(e.g. MB).  In other words, the structure function
at a particular lag is the normalised variance of the
difference pdf corresponding to that lag.  It therefore
represents only part of the information available from
the full pdf, which we consider below.

Grey-scale images of the centroid velocity differences for each
region are presented in Figures \ref{dvO} and \ref{dvM}, for a
fixed vector lag, in pixels, of $\ttau = [1,1]$.  In other words,
the images of Figures \ref{dvO} and \ref{dvM} were obtained by
displacing the maps of Figures \ref{vcO} and \ref{vcM}
horizontally and vertically 1 pixel and then computing the
difference as expressed by equation (\ref{DeltaV}).
Of particular interest when considering the centroid velocity
differences is the spatial distribution of large-amplitude
velocity differences--those events that compose the pdf tails.
Such events have been emphasised in Figures \ref{dvO} and \ref{dvM}
by using a 3-level grey scale.
As indicated by the intensity table in Figure \ref{dvM}, all
pixels in which the magnitude of the velocity difference is
more than twice the standard deviation of the map are shown
as white.  Magnitude values between one and two standard
deviations are shown as light grey, while values less than
one standard deviation are shown as dark grey.

It should be noted that many of the extreme events found in Figures
\ref{dvO} and \ref{dvM} are likely spurious, influenced by low
signal-to-noise ratios near the cloud edges and by residual
emission from overlapping cloud components (see Appendix \ref{noise}).  
This can be seen especially in Orion B, region 1b, L1551, and in 
the easternmost portions of Orion B, regions 2 and 4.  However, 
the distribution of extreme events within the main volume of 
the clouds is more reliable and generally reveals a very intermittent 
structure.  In most cases, high-amplitude velocity differences appear 
to be well distributed throughout the clouds, although there is also
some indication that some such events are correlated with the dense,
star-forming cores of Orion B, region 4 and Mon R2, region 1
(see also Figs. \ref{vimO} and \ref{vimM}).
At larger lags, the distribution of large-amplitude events
becomes less intermittent, as demonstrated in Figure \ref{dvO4}
for Orion B, region 4.

An important question is whether or not the largest velocity differences 
occur in correlated, filamentary structures, as in the simulations of 
mildly supersonic, decaying turbulence discussed by \cite{lis96} (1996).  
We see little evidence for such structures in the images of Figures \ref{dvO} 
and \ref{dvM}, which appear to have a relatively more ``spotty'' spatial
distribution.  Although noise-induced fluctuations near low-intensity 
cloud edges probably do contribute, the relatively spotty appearance persists 
even in the cloud interiors and near those cloud edges (e.g. the intensity 
ridge in the western portion of Orion B, region 4) where the intensity
is high and the influence of noise is small (see Appendix \ref{noise}).  

The qualitatively different spatial distribution of these observed velocity 
differences is likely due to the fact that the regions discussed here are 
highly supersonic and are continually driven by stellar energy input.  
In this case the velocity field is expected to be dominated by the 
compressional (shock) modes, not the vortical modes that dominated the
simulations discussed by Lis et al.  It is important to recognizee that,
except perhaps for scales smaller than about 0.1 pc, the
turbulent velocity field of the ISM should
be highly supersonic, and driven by shock interactions 
(\cite{kornr98} 1998).  Our results for the velocity
differences, along with the velocity pdfs themselves, as discussed in \S4.1
above, therefore suggest a fundamental difference between interstellar 
turbulence and incompressible turbulence, except perhaps in regions that 
have avoided energy input for a time long enough that they approach
incompressible conditions by means of turbulent decay.

However \cite{mivil98} (1998) report, based on
their study of HI in the Ursa Major cirrus
cloud, that the spatial positions corresponding to the non-Gaussian tails
of the difference pdf are primarily located
along three filaments in the cloud, although differences images
like our Figures \ref{dvO} and \ref{dvM} are not shown.
Such an image of velocity differences {\it is} presented in the CO (2--1)
study of the $\rho$Oph cloud by \cite{lis98a} (1998).  
Since the $\rho$Oph core is a region with active internal star
formation, it should be more directly comparable
with the regions studied in the present paper.  The image of large velocity
differences (Fig. 8 in \cite{lis98a} 1998) does
indicate that some of the highest velocity differences are filamentary,
although others are not.  Thus the situation is
not clear-cut, and the most we can claim is that, for the regions studied
here, there is little evidence for filamentary
clustering of regions with large velocity differences.
Quantification of this statement is beyond the scope of this paper, since 
establishing the reality or absence of filamentary structure is a difficult 
and unsolved problem, as is well-known from studies of the large-scale 
distribution of galaxies.

As in \S\ref{analysis}, we can investigate the velocity
differences further by considering their probability
density functions.  In order to facilitate visualisation
and improve statistics, we combine data from all lag
angles to obtain pdfs as a function of the lag magnitude,
$\tau \equiv \vert \ttau \vert$.  For each scalar lag,
$\tau$, measured in pixels, there is an associated pdf
which describes the distribution of velocity difference
values.  The pdf variation with $\tau$ is illustrated
in Figures \ref{dif_pdfsO} and \ref{dif_pdfsM}.  For
each region, solid, dashed, and dotted lines correspond
to scalar lags ($\tau$) of 1, 5, and 20 pixels.
All pdfs have been computed using the adaptive kernel
estimator described in \S\ref{analysis}.

Some of the difference pdfs in Figs. \ref{dif_pdfsO} and \ref{dif_pdfsM} appear 
to change from nearly-exponential at small lags to more nearly-Gaussian at
large lags, a behaviour which is similar to incompressible turbulence, 
the mildly supersonic decay simulations of \cite{lis96} (1996), and
the observational results of \cite{lis98a} (1998) and 
\cite{mivil98} (1998).  However many of the regions show a
persistence of exponential behaviour even at large lags ($|\tau |=20$
pixels), although sometimes with a cutoff at very
large velocity differences.  This persistence of exponential behaviour at
large lags is seen in most of the observed
regions, and again points to a difference between interstellar turbulence
and incompressible turbulence.  Alternatively,
the absence of a clear transition to Gaussian behaviour may simply reflect 
the fact that, at the largest lags examined, the
velocity field is still correlated while such larger-scale correlations do
not occur in the incompressible regime or in
the regions examined by \cite{lis98a} (1998) and \cite{mivil98} (1998).

\begin{figure*}[t]
\vspace{1in}\figcaption{Probability density functions of centroid velocity differences
for the six
subregions in Orion B.  In each plot, solid lines correspond to a scalar
lag, $\tau$,
of 1 pixel, while dashed and dotted lines correspond to scalar lags of 5 and 20
pixels respectively.  All pdfs shown are computed using the adaptive kernel
estimator
described in \S\ref{analysis}.\label{dif_pdfsO}}
\vspace{1in}\figcaption{Similar to Figure \ref{dif_pdfsO}, but for the Mon R2, L1228,
L1551,
and HH83 data sets.  Solid, dashed, and dotted lines again correspond to
the velocity difference pdfs at scalar lags of 1, 5, and 20
pixels.\label{dif_pdfsM}}
\end{figure*}

\begin{figure*}[b]
\vspace{1in}\figcaption{The variance of the difference pdfs is shown as a function
of scalar lag, $\tau$, measured in parsecs.  Frame (a) is comprised 
of the six subregions of Orion B and frame (b) includes the three
subregions of Mon R2 and the remaining L1228, L1551, and HH83
data sets.  The assumed distances to each region are as listed
in Table \ref{obs}. A power law of index 0.85 is shown in each
plot for comparison.\label{ad_var}}
\vspace{1in}\figcaption{Similar to Figure \ref{ad_var}, but here the kurtosis of the
difference
pdfs is plotted as a function of scalar lag, $\tau$.  As indicated by the
legend
in Figure \ref{ad_var}, frame (a) corresponds to Orion B, and frame (b)
corresponds to Mon R2, L1228, L1551, and HH83.  A power law of index
-0.5 is shown in each plot for comparison.\label{ad_kurt}}
\end{figure*}

The increase of the width of the velocity difference pdf with increasing
lag is qualitatively similar to
results found for incompressible turbulence.  However the same behaviour may
be expected for any stochastic
field:  at larger separations, the variables become less correlated, and
hence a larger fraction of the
differences have large values.  Basically this result only shows that the
structure function (variance of
velocity difference) increases with increasing lag, a property shared by a
large class of stochastic processes
(e.g. Burgers turbulence, which is entirely compressible).  In order to
constrain models for interstellar
turbulence, it is the functional form of this increase, as well as the
behaviour of other moments, which is crucial.

We therefore consider the variation with lag of the central
sample moments of the difference pdfs, defined as in
equations (\ref{M1})--(\ref{M4}).
The second moment, or variance, for each region is presented
in Figure \ref{ad_var}, as a function of the scalar lag,
$\tau$, expressed in parsecs.  As mentioned above, the
variance of the difference pdfs is simply the structure
function, without the customary normalisation factor.
The six subregions of Orion B are shown in frame
(a), and frame (b) shows the three subregions of
Mon R2 together with the L1228, L1551, and HH83 results.
The variation of the normalised fourth moment, or kurtosis
[eqn. (\ref{M4})], with scalar lag for each region is displayed
in Figure \ref{ad_kurt}.

It is clear from Figures \ref{ad_var} and \ref{ad_kurt} that the variation
of the difference moments over some lag ranges can be approximately described
by a power-law dependence, which appears linear on the log-log axes shown.
A characteristic power-law index can be derived for each region by measuring
the mean logarithmic slope of each curve over some chosen range in lag.
However, the choice of the fitting range is not straightforward--most of
the curves in Figures \ref{ad_var} and \ref{ad_kurt} exhibit steep slopes
at small lags which flatten out as the lag increases.  This ``flattening
out'' is a common feature of all stochastic fields, which approach a constant
variance and kurtosis at lags far exceeding the field's characteristic
correlation length, or as in the present context, the effective width of
the applied spatial filter. The fitting range should therefore not include
large lags, where the slope is systematically small.  On the other hand,
sampling effects can influence the slope at the smallest lags
(MB).  We therefore chose the intermediate range of lags between 6 and 12 
pixels (note that this corresponds to a different physical scale in each
region), and applied power-law fits of the form:
\be{momexp}
\mbox{variance} \propto \tau^{\alpha_2} \mbox{\hspace{.5in} and \hspace{.5in}}
\mbox{kurtosis} \propto \tau^{\alpha_4} \; .
\ee
The results yielded $\alpha_2$ values between 0.64 and 1.05 (except for
Orion B, region 1b, in which $\alpha_2=0.33$) and $\alpha_4$ values
between -0.11 and -0.88, with typical values of 0.85 and 
-0.5 respectively.    These typical values are shown on the plots
of Figures \ref{ad_var} and \ref{ad_kurt} for comparison with the
data.  We emphasise that these best-fit power law exponents 
depend significantly on the chosen fit range and should therefore 
be viewed only as a rough estimate of the logarithmic slope of 
the moment curves at intermediate lags.  With the exception of
the L1228 variance ($\alpha_2$ = 0.66), the smaller-scale, 
low-mass star-forming regions L1228, L1551, and HH83 exhibit 
somewhat steeper logarithmic slopes 
(1.01 $\leq \alpha_2 \leq$ 1.05, -0.88$\leq \alpha_4 \leq$ -0.64)
than the GMC regions (0.33 $\leq \alpha_2 \leq$ 0.90, 
-0.52 $\leq \alpha_4 \leq$ -0.11), possibly reflecting the influence 
of molecular outflows.  The smallest slopes occur in the
``cometary clouds'', regions 1a and 1b of Orion B, although it
should be noted that these regions have the poorest statistics
(in terms of both signal-to-noise and number of spectra).

The variance results reported here are consistent with the more
detailed structure function analysis of MB.  The reader
is referred to that paper for further work and for a comparison with
structure function indices predicted by various analytic and
numerical turbulence models and with empirical scaling relations
from interstellar observations (e.g. ``Larson'' scaling) and
laboratory turbulence experiments (e.g. ``Kolmogorov'' scaling).

The variation of kurtosis (flatness) with lag can be compared with the
incompressible turbulence experiments of \cite{vanat80} (1980) 
and the multifractal models of \cite{egger98} (1998),
which exhibit a basically constant kurtosis out
to some critical lag (which depends on the Reynolds number), beyond which
the kurtosis decreases with increasing lag.
However, except for a narrow transition region near the critical lag,
this decrease is much slower than that found in the present work.

We believe that the ability of hydrodynamical simulations to account for the
observed variations of second (Fig. \ref{ad_var}) and fourth (Fig. \ref{ad_kurt}) 
moments of the velocity difference pdfs with lag provides a decisive test of 
turbulence models, although it must be remembered that the observations refer to
regions that are supersonic and contain internal turbulent power sources.

\subsection{Reynolds Numbers}\label{Resec}
The dynamical information available in the centroid velocity and velocity
difference maps considered above can be used to assess the ``degree'' of
turbulence in each region, as quantified by the Reynolds number.
The Reynolds number $Re=v\ell/\nu$ is defined only with respect to the scale
$\ell$,
which might be the size of the region, the correlation length (integral
scale), or any other length that measures a
useful characteristic spatial scale of the turbulence.  One such scale
which has proven especially useful in
characterising incompressible turbulence experiments and simulations is the
Taylor microscale, which measures the
average spatial extent of velocity gradients.  For incompressible
turbulence, these gradients are entirely due to
the vorticity, so the Taylor microscale is defined as 
(e.g. \cite{lesie90} 1990, p. 144)
\be{Reg}
\ell_T = \frac{u_{rms}}{\langle({\bf \nabla}\times {\bf u})^2\rangle^{1/2}}
\ee
where $u_{rms}$ is the root-mean-square fluid velocity.  For compressible
turbulence the velocity gradients are due
to a combination of vortical (rotational, solenoidal) and compressible
(irrotational, dilatational) modes.  For
highly supersonic turbulence the vorticity in eqn. (\ref{Reg}) might be replaced by
the dilatation
${\bf\nabla}\cdot{\bf u}$.  (In this case the simple scaling between Taylor
scale Reynolds number and integral
scale Reynolds number [eq. VI-6-7 in \cite{lesie90} 1990, p. 144] is lost because
there are no relations between enstrophy
and viscous dissipation rate and the dissipation rate cannot be taken as
$u^3/\ell$ as for the dissipationless
Kolmogorov energy cascade.)

\begin{deluxetable}{lccccccccc}
\small
\tablecaption{Reynolds Numbers and Taylor Scales\label{ReTa}}
\tablehead{\colhead{Region} & \colhead{L} & \colhead{n}
 & \colhead{$\sigma_c$} & \colhead{s} &
\colhead{$\sigma_s$} &\colhead{Re$_L/10^7$} & 
\colhead{Re$_T/10^6$} & \colhead{$\ell_T$} \\
& \colhead{(pc)} & \colhead{(10$^2$cm$^{-3}$)}
 & \colhead{(km s$^{-1})$} & \colhead{(pc)} &
\colhead{(km s$^{-1}$)} & & & \colhead{(pc)} } 
\startdata
Orion B (1a) & 6.0 &2 &0.55 &0.12 &0.10 &1.2 &1.3 &0.66 \nl
\phm{Orion B }(1b) & 8.6 &2 &0.55 &0.12 &0.19 &1.7 &0.69 &0.35 \nl
\phm{Orion B }(1c) & 17 &2 &0.83 &0.12 &0.21 &5.1 &1.4 &0.47 \nl
\phm{Orion B }(2) & 30 &2 &1.15 &0.12 &0.21 &12 &2.7 &0.66 \nl
\phm{Orion B }(3) & 33 &2 &2.05 &0.12 &0.41 &24 &4.4 &0.60 \nl
\phm{Orion B }(4) & 40 &2 &1.31 &0.12 &0.25 &19 &3.0 &0.63 \nl
\tablevspace{.1in}
Mon R2 (1) & 41 &6 &1.12 &0.24 &0.19 &50 &17 &1.4 \nl
\phm{Mon R2 }(2) & 32 &6 &0.74 &0.24 &0.12 &26 &12 &1.5 \nl
\phm{Mon R2 }(3) & 50 &6 &0.83 &0.24 &0.16 &45 &11 &1.2 \nl
\tablevspace{.1in}
L1228 & 2.6 &10 &0.74 &0.087 &0.064 &3.5 &13 &0.59 \nl
L1551 & 1.8 &10 &0.50 &0.041 &0.085 &1.6 &2.2 &0.24 \nl
HH83 & 0.67 &25 &0.42 &0.014 &0.042 &1.2 &2.6 &0.14
\enddata
\end{deluxetable}

The Taylor scale Reynolds number is generally smaller than the integral
scale Reynolds number.  For reference,
simulations of incompressible turbulence are only capable of generating
turbulence with Taylor scale Reynolds
numbers of at most 10$^2-10^3$, while experiments reach Taylor scale
Reynolds number of at most $\sim10^3-10^4$
(see, for example, \cite{kaila92} 1992, \cite{she93} 1993, and
references therein).

For empirical studies such as the present one, it is appropriate to use the
rms value of the measured velocity
gradient in the denominator of eqn. (\ref{Reg}).  This is
clearly an important quantity, since it gives a measure of the
characteristic scale over which turbulent interactions of any sort (e.g.
shocks, vorticity stretching) occur.  A
problem is that estimation of centroid velocity gradients across the line
of sight will be strongly amplified by
uncertainties in the centroid velocities, while if we try to avoid this
problem by smearing the centroid velocity
field with a filter, the resulting rms velocity gradient will decrease with
increasing filter size.  A convenient
way to circumvent these problems is to take the appropriate mean velocity
gradient as the square root of the
variance of the velocity difference pdf taken at the smallest separation
(one pixel), which we will call $\sigma_s$, divided by the physical length 
corresponding to this separation, which we will call $s$.  The reasoning 
is that the average taken when computing the variance of the
difference pdf will greatly reduce the noise due to uncertainties in
individual centroid velocities.  We therefore
define the Taylor scale as
\be{ellT}
\ell_T=\frac{\sigma_c \; s}{\sigma_s}
\ee
where $\sigma_c$ is the rms centroid velocity fluctuation amplitude.  
Taking the velocity occurring in the Reynolds number as
$\sigma_c$, this gives a Taylor scale Reynolds number
\be{ReT1}
Re_T=\frac{\sigma^2_c \; s}{\nu\sigma_s}
\ee
We take $\sigma_c$ to be the dispersion of the mean spectral line
profile for each region, given by the ``parent dispersion''
$\sigma_p$ in Table 2 of MB.  Alternatively,
the ``turbulent dispersion'', $\sigma_t$ in Table 2 of MB, 
could be used, but it includes uncertainties due
to instrumental noise and it removes the influence of large-scale
velocity variations under the assumption that they are uncorrelated
with small-scale fluctuations.  Since large scales in turbulent
flows typically contain much of the kinetic energy and
play an important role in the nonlinear energy transfer among
modes, $\sigma_p$ is the most appropriate dispersion measure
to use in computing an effective Reynolds number.
However, the difference is relatively minor.

Note that our definition of the Taylor scale (eqn. \ref{ellT}) and 
the associated Reynolds number (eqn. \ref{ReT1}) depend on the scale
at which they are computed--i.e. the spatial resolution of the map.
However, this is reasonable because it is the scale of the fluid
motions we are sampling - we have no information about the dynamics
on scales smaller than the resolution or larger than the emission
region.  The values may change if higher resolution maps become
available, but only if there is a large amount of power at high
spatial frequencies.  The range of values found for the Taylor
scale, 3 to 10 times larger than the grid spacing, suggests 
that this may not be the case.  Projection effects and
line-of-sight averaging also influence the computation of the Taylor 
microscale, but these effects are very difficult to quantify without 
some assumption about the three-dimensional nature of the velocity field.
They could lead to an overestimate of the Taylor microscale and 
the corresponding Reynolds number, although to our knowledge
this has never been demonstrated.

The molecular viscosity is estimated using a mean free path approximation,
$\nu=v_{th}/\sigma n$, where $v_{th}$ is the rms
thermal velocity (kT/$\mu$m$_H)^{1/2}$, $n$ is the particle number density,
and $\sigma$ is the collisional cross
section.  We adopt the cross section for $H_2-H_2$ elastic collisions as
about 10$^{-15}$ cm$^2$. It is unclear to us whether the absence
of a dipole moment for the H$_2$ molecular
should alter the elastic cross section compared to typical values for other
atoms and molecules reported in the
literature; we assume it does not.  The particular densities
are taken as the rough estimates given by MB: n(cm$^{-3}$) = 2500 for HH83,
1000 for L1228 and L1551, 600 for Mon R2, and 200 for the Orion B subregions.
We adopted a characteristic temperature
of 20 K for all the regions.

Then the Taylor scale Reynolds number can be expressed as
\be{ReT}
Re_T=1.8\times10^6 n_2\sigma_{c,5}^2 s_{pc}/\sigma_{s,5}
\ee
where n$_2$=n/100 cm$^{-3}$, $\sigma_{c,5}$ is the turbulent velocity
dispersion in km s$^{-1}$, $s_{pc}$ is the
linear scale corresponding to a one pixel lag at the adopted distance of
the region, and $\sigma_{s,5}$ is the
standard deviation of the velocity difference pdf at this lag, in km
s$^{-1}$.  For comparison, the Reynolds number
corresponding to the scale L of the observation, taken as the 
geometrical mean of the semiminor and semimajor axis of
the emitting region on the plane of the sky, is 
\be{ReL}
Re_L=1.8\times10^6 n_2\sigma_{c,5}L_{pc} \; .
\ee

We computed these dimensionless numbers for each region and 
the results are given in Table \ref{ReTa}.  The region scale Reynolds
number $Re_L$ is, as expected, very large, varying between about
1$\times10^7$ to 5$\times10^8$.  This range is very nearly the
same as the estimates of $Re_L$ given by \cite{myers95} (1995) 
for a sample of ``diffuse,'' ``dark,'' and ``giant'' clouds.
 Surprisingly, the Taylor scale Reynolds numbers, while generally
smaller than
$Re_L$, are still very large, between 7$\times10^5$ and
2$\times10^7$. As mentioned above, the Taylor microscales, which 
measure the rms scale of velocity gradients, are, in units of the
resolution s, $\sigma_c/\sigma_s=$3 to 10, suggesting that we 
are resolving the rms gradient scale.  However, the very large values 
of the Taylor scale Reynolds number mean that, at the smallest 
resolvable scales, the advection term in the momentum equation, as
measured by the velocity gradients, is still huge compared to viscous 
dissipation, implying that we have not yet resolved the dissipation
scale, so velocity gradients must still be present on spatial scales
smaller than the sampling grid of the observations.
Considering the numerical diffusion/artificial viscosity in existing
numerical codes, simulations are far from approaching the resolution
necessary to realistically represent interstellar turbulence.
However, in the presence of a magnetic field, the dissipation might 
be due to ion-neutral friction or magnetic reconnection rather 
than ordinary viscosity (\cite{myers95} 1995), making the situation 
less severe.

\placetable{ReTa}

     A surprising result apparent in Table \ref{ReTa} is that the Taylor scale 
$\ell_T$ is approximately constant among most of the subregions of Orion B 
($\ell_T \approx$ 0.6 pc) and of Mon R2 
($\ell_T \approx$ 1.4 pc).  The difference between Orion B and Mon R2 might be
attributed to the differing spatial resolutions, since 
$\ell_T$ (eqn. \ref{ellT}) is
proportional to the size of a resolution element, and Mon R2 is roughly
twice as distant as Orion B.  The same distance scaling is apparent for L1551,
and qualitatively for L1228 and HH83, where $\ell_T$ is only 0.14 pc: 
HH83 is at about the same distance as Orion B, but was observed at about 
eight times better resolution.  These results may suggest that the
Taylor microscale is approximately constant in all the regions observed
when the reference spatial scale $s$ (eqn. \ref{ellT}) is normalised to the 
same value.  This is a particularly intriguing result, since it would 
imply that the characteristic rms scale of velocity gradients, measured relative 
to the rms global velocity field, is a characteristic length scale independent 
of differing physical conditions in the regions studied.  However, we are
unable to assign a value to this scale because it depends on the adopted
scale over which the gradients are measured, taken here as the size of a
resolution element.

Experiments (\cite{vanat80} 1980; \cite{tabel96} 1996) and
multifractal models (\cite{bifer93} 1993; \cite{egger98} 1998)
of incompressible turbulence indicate that the kurtosis (flatness) of the
velocity difference pdf should increase with
$Re_L$ as about $R_L^{0.15}$, but for Taylor scale Reynolds numbers greater
than about 700 the kurtosis decreases with
$Re_L$.  We find no evidence for such behaviour in the observations
presented here, again suggesting that highly
compressible turbulence differs significantly for incompressible
turbulence.  However it should be noted that both
$Re_L$ and $Re_T$ are proportional to adopted average densities, which are
very uncertain, and so errors in the density
estimates may mask any correlations.

We have not attempted to estimate magnetic Reynolds number 
(see \cite{myers95} 1995) for each region, because the
magnetic viscosity depends on the magnetic field strength and the ionized
fraction, quantities which are unknown for the
present regions.  The significance of these numbers for the propagation of
MHD waves is discussed in \cite{myers95} (1995).

\section{Summary}\label{summary}
	The probability distribution function (pdf) for fluctuations of
molecular line centroid velocities, and line centroid velocity differences 
at different separations, have been estimated for a number of local regions 
with active internal star formation.  The data consist of a total of over 
75,000 $^{13}$CO line profiles covering five different molecular clouds.
The internal stellar power sources include only low-mass protostellar winds in 
three regions (L1228, L1551, HH83) but extend to massive stars in
the remaining two regions (Orion B and Mon R2).  The GMCs Orion B and Mon R2
were subdivided into six and three kinematically distinct subregions
respectively in order to isolate particular cloud components. 
The total sample therefore numbers twelve distinct regions, each
composed of about 1000 to 25000 independent spectra.  Centroid velocity 
fluctuation maps were constructed by interpolating these spectra onto regular 
spatial grids and then applying spatial filters in order to remove large-scale 
gradients.  These maps are displayed as images in Figs. \ref{vcO} and \ref{vcM}.

	The pdf of centroid velocities for each region was estimated using
the classical histogram, a non-parametric adaptive kernel estimator, and
the parametric Johnson estimator.  The methods generally
agree well except in the far tails of the pdfs.  Although the Mon R2 regions 
exhibit nearly-Gaussian pdfs, except possibly in the far tails,
all the other regions show strong excesses relative 
to a Gaussian, often suggesting nearly-exponential or power-law forms.  
Sample moments and stretched exponential fits to the pdfs were presented as 
quantitative measures of the departure from Gaussian statistics.
These results confirm and extend the general conclusions reached by MS.

	Centroid velocity pdfs for diffuse interstellar HI regions,
constructed from older published
optical line and HI emission and absorption line
data, are also presented, and also show strong evidence for non-Gaussian,
nearly exponential, pdfs.  The similarity with most of the molecular
regions studied here suggests that either the exponential pdfs are not a
product of stellar activity or that the internal velocity fields of the
diffuse regions have been strongly affected by disturbances from external
sources.  These exponential centroid pdfs do not agree with the
nearly-Gaussian pdfs found in numerical simulations of freely-decaying
marginally supersonic turbulence by \cite{lis96} (1996).  The pdfs are also
markedly different from the nearly Gaussian velocity pdfs found in studies
of incompressible turbulence.  A theoretical interpretation of the centroid
pdfs is presented in a separate paper (Paper III).

	We also constructed the pdfs of centroid velocity differences for
lines of sight separated by different scalar spatial lags.  In agreement
with other recent observational work on this function, and with studies of
incompressible turbulence, we find nearly exponential difference pdfs 
at small lags which broaden with increasing lag.
However, spatial images of the centroid
differences show a ``spotty'' distribution for the largest velocity
differences, with little evidence for the filamentary structures predicted
by simulations of decaying marginally supersonic
turbulence--structures which should be due to vorticity.

	We used the velocity difference pdfs to estimate the Taylor
microscale, which is the rms scale of velocity gradients.  The Taylor
microscale Reynolds number for all the regions is very large, $\sim$10$^5$
to 10$^6$, indicating that even at the smallest resolvable scales, the
advection term in the momentum equation is still huge compared to viscous
dissipation.

Our observational results suggest fundamental differences between
turbulence in both star-forming regions and diffuse atomic interstellar 
clouds as compared with incompressible or mildly-supersonic decaying 
turbulence.  These interstellar gas motions are characterised 
by very large Reynolds numbers, a high degree of
compressibility, continuous energy and momentum injection by internal and external 
power sources, and various sources of anisotropy, including rotation, shear, 
magnetic fields and self-gravity.  It might be necessary to incorporate some or 
all of these characteristics into numerical and theoretical models before full 
agreement with observations can be achieved. The complexity of interstellar 
turbulence suggests that it may exhibit some degree of nonlinear self-organisation, 
producing coherent structures which could contribute non-Gaussian components to the 
velocity and velocity difference pdfs.  Such behaviour has recently been demonstrated 
in three-dimensional numerical simulations of turbulent incompressible shear 
flow (\cite{pumir96} 1996; \cite{lamba97} 1997) and turbulent compressible convection 
(\cite{brand96} 1996).  
On the other hand, it may be that some rather simple and fundamental property of the 
ISM is behind its statistical behaviour.  For example, stellar winds, outflows, 
and superbubbles can all produce non-Gaussian velocity distributions with power-law
pdf tails (e.g. \cite{silk95} 1995; \cite{oey98} 1998).  Alternatively, as another example, 
the ISM is a system in which kinetic energy is not a global invariant conserved by
the advection operator as it is for incompressible turbulence; in such a situation 
it is possible to obtain exponential velocity distributions, as arise in simulations 
of systems of interacting wind-driven shells (\cite{chapp99} 1999).  In Paper III, 
we will further address these and related issues regarding the theoretical implications 
and interpretation of the results reported in the present paper.

This work was supported by NASA grant 5-3107 to JS, and by the National Research
Council, through a postdoctoral fellowship to MM.

\appendix
\section{Appendix A: Overview of Regions Studied}\label{overview}
All of the regions considered here are actively forming stars
and there is evidence to suggest that the stars which have
been produced have played a significant role on the dynamics
of their parent clouds.  In this appendix we provide a brief
description of the environments and physical conditions in
each of the regions studied in order to aid in the physical
interpretation of the statistical analysis of \S\ref{results}

\subsection{Orion B}\label{env_ori}
The Orion giant molecular cloud (GMC) complex is one of the nearest
and most popular sites for studying massive star formation
and its interaction with surrounding interstellar material.
The GMC is comprised primarily of the A cloud (L1641), which lies
behind the Trapezium stellar cluster and the well-known Orion nebula,
and the B cloud (L1630), which is of a comparable size and mass and
extends out to the northeast.  The molecular gas in the region has been
mapped extensively in $^{12}$CO by \cite{madda86} (1986), in the 
associated dust emission using IRAS (\cite{beich88} 1988; 
\cite{robin84} 1984; \cite{wood94} 1994),
and in other isotopes of CO and CS (e.g. \cite{bally91} 1991; \cite{lada91} 1991;
\cite{krame96} 1996).  The star formation activity and associated cloud
dynamics have been reviewed by, e.g., \cite{genze89} (1989).

The Orion GMC complex may have originated through the fragmentation and
local gravitational collapse of the expanding Gould's Belt supershell
about 10 to 20 Myr ago \cite{bally95b} (1995) or from the collision of an
infalling high-latitude cloud with the galactic plane (\cite{franc88} 1988).
Massive star formation began by 12 Myr ago and produced the I Orion OB
association.  There is good evidence that these young, hot stars
have had a profound influence on their parent clouds by accelerating,
ionizing, ablating, compressing, and disrupting them through supernovae,
radiation, and stellar winds (\cite{bally87} 1987; \cite{bally89} 1989; 
\cite{bally91} 1991; \cite{genze89} 1989).  The large-scale velocity gradients and 
morphologies of the main Orion A and B clouds as well as the ``wind-swept'' or 
``cometary'' appearance of a number of nearby  smaller clouds, with ``tails'' 
directed away from the OB association (as in regions 1a and 1b in the present 
study), provide a strong indication that the association has played an important
role in their dynamics.

The Orion B mapping considered here (Fig. \ref{vimO}) covers a number of
regions
which are still actively forming stars.  The reflection nebulae NGC 2068
and 2071 are located near emission peaks in what we have defined
as region 2, and the main peaks along the emission ridge in region
4 are near the well-known HII regions NGC 2023 and 2024 and the
Horsehead Nebula B33.  Other notable objects within the mapped
region include a large number of reflection nebulae, embedded infrared
sources, HII regions, Herbig-Haro objects, H$_2$O masers, and molecular
outflows (see \cite{madda86} 1986, \cite{reipu85} 1985, \cite{lada92} 1992,
\cite{chand96} 1996 and references therein).  In an infrared survey
of the area, \cite{lada91b} (1991) identified approximately 1000
sources, about half of which are probably embedded in the Orion B
cloud.  Most of these sources are clustered, with $\sim$ 330 near
NGC 2023 and 2024 in our region 4, and another $\sim$ 300 near NGC 2068
and 2071 in our region 2.  Active star formation is even occurring in
at least some of the smaller clouds in region 1, as demonstrated by
\cite{reipu91} (1991), who report several Herbig-Haro jets and molecular
outflows in L1617 (which we have labelled region 1b).

In summary, the Orion B observations studied here represent a very
dynamic region where the molecular gas is substantially influenced
by supernovae, radiation, and winds from the massive stars in the
nearby OB association and from embedded young stars which are
continually forming within.

\subsection{Mon R2}\label{env_mon}
The Monoceros R2 molecular cloud is comparable in size and
CO luminosity to Orion B.  The large-scale distribution 
of $^{12}$CO follows well the area spanned by the Lynds dark clouds 
L1643, L1644, L1645, and L1646 (\cite{madda86} 1986).

The core of the Mon R2 cloud lies at the origin of the coordinate
system used in Figure \ref{vimM}a, near the predominant CO emission peak,
and is a well-studied region of ongoing high-mass star formation.  
It has been extensively mapped in CO isotopes, CS, HCN, H$_2$CO, NH$_3$,
and HCO$^+$, and is associated with a number of B stars and
reflection nebulae, a large cluster of embedded infrared
sources, a compact HII region, enhanced X-ray emission, and both
H$_2$O and OH masers (see \cite{beck76} 1976, \cite{thron80} 1980,
\cite{monta90} 1990, \cite{torre90} 1990, \cite{gonat92} 1992,
\cite{giann97} 1997, \cite{tafal97} 1997, \cite{grego98} 1998,
and references therein).  The embedded cluster has recently been
studied in detail by \cite{carpe97} (1997), who estimate that its
population numbers at least 475 stars.  Also associated with
this core region is one of the largest, most massive molecular
outflows known, extending for at least 4 parsecs along its axis
and involving almost 200 $\Msun$ of interstellar material
(\cite{wolf90} 1990; \cite{meyer91} 1991; \cite{xie93} 1993).  The outflow
is bipolar, centered near the middle of the embedded cluster, and
dominates the dynamics of the molecular cloud within at least
several arcminutes of the core (corresponding to the origin in 
Figure \ref{vimM}a).  A separate outflow, located about 75$^{\prime\prime}$ 
from the main embedded cluster, was recently identified 
by \cite{tafal97} (1997).

A second prominent CO emission peak lies about 45$^{\prime}$ to the
east (in region 2), and corresponds to the molecular core
known as GGD12--15 (\cite{littl90} 1990).  Evidence indicates that this
region too is actively forming stars and is also associated with
a large bipolar molecular outflow, a compact HII region, several
infrared sources, and an H$_2$O maser (\cite{rodri80} 1980;
\cite{olofs85} 1985; \cite{littl90} 1990).  Furthermore, there are a
number of IRAS point sources tracing the sharp emission ridges in
regions 2 and 3 as well as additional infrared sources and
reflection nebulae scattered primarily throughout regions
1 and 2 and another HII region near the northeastern edge of
region 1 (\cite{xie94} 1994).  At least 30 to 40 separate infrared
and X-ray sources have been identified outside the main
Mon R2 core (\cite{xie94} 1994; \cite{grego98} 1998).

\cite{loren77} (1977) studied the large-scale kinematics of the region
using CO, H$_2$CO, and near-infrared observations and he interpreted
velocity gradients across the cloud as a combination of rotation
and collapse.  However, the more recent, higher-resolution mappings
by \cite{xie94} (1994) have revealed a more complicated intensity and velocity
structure, characterised by sharp emission ridges in the western and
northeastern portions of the cloud and a southeastern portion with
a lower LSR velocity and a more ``wispy'' appearance.  This overall
structure is confirmed by the observations presented here and is the basis
for the region decomposition described in \S\ref{data}.  \cite{xie94} (1994)
interpreted the relative blueshift of region 1 as evidence for a large
expanding shell, involving $\sim$ 4 $\times 10^4 \Msun$ of material and
moving toward us with a velocity of $\sim$ 3--4 \kms.  Fairly recent,
large-scale dynamical events are also suggested by the sharpness of
the emission ridges in regions 2 and 3 and the evidence for associated
star formation (see the preceding paragraph), which may indicate shock
compression of the molecular gas.

In summary, Mon R2 is a giant molecular cloud comparable in scale to Orion B.
Although there is no analogous OB association disrupting the molecular gas
to the same degree as in Orion, active star formation is indeed occurring
within Mon R2 and there is also some evidence suggesting violent dynamical
events on the scale of the cloud.

\subsection{L1228 and L1551}\label{env_lynds}
The Lynds dark clouds L1228 and L1551 are best known for the
molecular outflows found within them.  Both regions are nearby,
well-studied sites of ongoing low-mass star formation.

The L1228 cloud is located in the Cepheus flare, in a ring of
molecular gas which seems to be part of a 4$\times 10^4$ year
old supernova remnant (\cite{greni89} 1989;
\cite{yonek97} 1997).  At the center of the coordinate
system in Figure \ref{vimM}b lies the young stellar object
IRAS 20582+7724 and associated with it, extending
northeast and southwest, is a large (18$^\prime$ $\times$
9$^\prime$), well-collimated, bipolar molecular outflow
(\cite{haika89} 1989; \cite{bally95} 1995; \cite{angla97} 1997;
\cite{tafal97b} 1997).  The orientation and sense of the outflow
axis is roughly the same as the large-scale velocity gradient
apparent in Figure \ref{vimM}b.  Also, \cite{bally95} (1995) have 
argued that there is at least one other YSO producing an outflow in the
L1228 cloud core and have reported a number of HH objects in the
vicinity.  Further evidence for active low-mass star formation
has been provided by \cite{ogura90} (1990), who identified
69 H$\alpha$ emission stars in the area, nine of which are
concentrated near the L1228 cloud core.

L1551 occupies a small area along the edge of the Taurus-Auriga
molecular cloud complex (e.g. \cite{unger87} 1987) and is one of the
most fertile centers of low-mass star formation known.  The large
(10$^\prime$ $\times$ 35$^\prime$), collimated,
bipolar molecular outflow  in the southeastern portion of the
cloud associated with the infrared source IRS 5 (which lies at
the origin of the coordinate system in Fig. \ref{vimM}c) is one of the
archetypical examples of the outflow phenomenon in young stellar
objects (YSOs) and has been the subject of much observational
and theoretical research (see \cite{snell80} 1980, \cite{cabri86} 1986,
\cite{uchid87} 1987, \cite{moria88} 1988, \cite{stock88} 1988,
\cite{bachi94} 1994, \cite{davis95} 1995, and references therein).
In the coordinate system chosen in Figure \ref{vimM}c, the outflow axis
is oriented approximately horizontally, with blue-shifted gas
toward the right of the origin and red-shifted gas toward the
left, and it occupies much of the lower portion of the map.
Also associated with IRS 5 and its accompanying outflow are
a large reflection nebula known as HH102 and several other
Herbig-Haro objects, including HH28 and HH29
(e.g. \cite{graha92} 1992).

In addition to the IRS 5 vicinity, there are several other regions
throughout the L1551 cloud which exhibit strong evidence for
ongoing low-mass star formation.  Especially several arcminutes
to the north, near the very young T Tauri stars HL and XZ Tau
and the YSO HH30, a large number of Herbig-Haro objects, molecular
outflows, optical jets, infrared sources, and excess H$\alpha$ emission
regions have been identified (\cite{mundt88} 1988; \cite{mundt90} 1990;
\cite{graha90} 1990; \cite{pound91} 1991).  Recent X-ray observations by
\cite{carkn96} (1996) and CCD and spectroscopic observations by 
\cite{brice98} (1998)
have increased the total number of known T Tauri stars in the L1551 cloud
to at least 26, along with one young B9 star.

\subsection{HH83}\label{env_hh83}
The structure, dynamics, and nature of the HH83 molecular cloud have been
described in detail by \cite{bally94} (1994).  The innermost core has also been
mapped in CS emission by \cite{nakan94} (1994).  This cloud is smaller than
the others in our study (see Table \ref{obs}), and is located at the extreme
western edge of the Orion A cloud (L1641), within an area to the northwest
of NGC1999 which has a high concentration of Herbig-Haro objects and several
CO outflows, signifying active star formation in the region.

An embedded star--the infrared source HH83 IR--lies in the southwestern
part of the cloud, at the origin of the coordinate system shown in
Figure \ref{vimM}d, and an associated optical jet extends out toward the
northwest
for at least 32$^{\prime\prime}$, terminating in a conical bow-shock about
150$^{\prime\prime}$ from the central IR source (e.g. \cite{reipu89} 1989).  Also
associated with the embedded star and the jet is a low-velocity ($\approx$
5 km s$^{-1}$), poorly collimated, bipolar outflow (\cite{bally94} 1994).
The outflow does appear in molecular emission lines, but there is good
evidence
that it is in a late stage of its evolution, having ``blown out'' of its
parent
cloud, and is currently depositing most of its energy and momentum into the
atomic inter-cloud medium.  Although the $^{13}$CO-emitting gas does not trace
the full extent of the outflow, small-scale velocity gradients near the
central
IR source and evidence for evacuated cavities along the jet axis suggest that
the outflow has had a substantial dynamical influence on the molecular gas in
the vicinity.

The axis of the outflow and jet, like several others in the region,
is approximately aligned with the mean, large-scale magnetic field
(directed northwest/southeast).  The cloud is elongated perpendicular to this
direction and, at least in the northern portion, exhibits a velocity gradient
which may indicate gravitational collapse along the mean magnetic field lines
on a timescale of $\sim$ 3$\times 10^5$ years.  Also, the large-scale gradient
across the cloud may indicate rotation about an axis aligned with the magnetic
field and outflow.  Furthermore, the cloud's proximity to the Orion OB
association to the north suggests that it may have been accelerated southward
relative to the surrounding gas by ablation, supernovae, and stellar winds.
The associated compression may have triggered gravitational collapse in the
southern end of the cloud, and may account for why star formation is currently
occurring in the southern portion, but seems to be absent in the northern
portion.

\section{Appendix B: Sources of Error}\label{noise}
The influence of instrumental noise on the centroid velocity fluctuations
was studied by \cite{miesc94a} (1994), and we only summarise their results 
here.  For each region studied in that (and in this) paper, they approximated 
the mean spectrum as a Gaussian with a peak, dispersion, and center determined 
respectively by the observed mean brightness temperature, linewidth, and 
LSR velocity.  The rms brightness temperature fluctuation due to instrumental 
noise was then estimated for each spectrum based on the effective integration 
times and detector temperatures, and a mean rms noise fluctuation was computed
for each data set.  A series of simulations was then performed for 
each region, based on the integration window used for the observations, 
the model Gaussian mean line profiles, and the mean noise fluctuations.
In the simulations, white noise of the appropriate rms amplitude was added to
the model profiles and the centroid velocity was computed.  This operation was
repeated (typically about 10000 times) with different random number seeds for
the white noise, and the variance of the centroid velocity in these simulations
was computed.  The results indicate that the noise-induced fluctuations are
very small, typically increasing the standard deviation of the centroid velocity 
by only a few percent (compare $\sigma_c^*$ to $\sigma_c$ in Table 2 of 
\cite{miesc94a} 1994).

Although these simulations do not directly address the influence of noise
on the far tails of the centroid velocity and centroid velocity difference 
pdfs, they do suggest that this influence is minor, except for those spectra 
near cloud edges where the brightness temperature drops substantially below 
the mean brightness temperature.  As discussed in \S\ref{data}, threshold
integrated intensity levels were used to eliminate spectra with very low
signal-to-noise, but some questionable spectra near cloud edges still remain.
For example, the extreme velocity and velocity difference values near the
edges of Orion B region 1b (Figs. \ref{vcO} and \ref{dvO}) and L1551 
(Figs. \ref{vcM} and \ref{dvM}) may be influenced by noise.

In some areas, the influence of kinematically distinct, spatially 
overlapping emission may also bias the centroid pdfs, although 
the velocity ranges were carefully chosen to minimise this effect
(see \S\ref{data}).  This is particularly the case for the eastern
portion of Orion B, region 4 (Figs. \ref{vcO} and \ref{dvO}).

\end{document}